\documentclass[thm-restate,numberwithinsect,a4paper,UKenglish]{lipics-v2021}

\usepackage{tikz}
\usetikzlibrary{calc,fit,decorations.pathreplacing}
\usepgflibrary{arrows}
\usepackage{rotating}
\usepackage{float}
\usetikzlibrary{positioning}
\usepackage{tikz-cd}

\hideLIPIcs
\nolinenumbers

\theoremstyle{definition}
\theoremstyle{remark}
\newcommand{\ML}{\text{ML}}
\newcommand{\PL}{\text{PL}}
\newcommand{\M}{\mathfrak{M}}

\newcommand{\F}{\mathfrak{F}}
\newcommand{\G}{\mathfrak{G}}
\newcommand{\A}{\mathfrak{A}}

\renewcommand{\phi}{\varphi}
\newcommand{\tr}{\mathit{tr}}
\newcommand{\rt}{\mathit{rt}}

\newcommand{\bisim}{\sim}

\newcommand{\inv}{\mathit{cover}}
\newcommand{\charf}{\mathit{char}}

\newcommand{\Prop}{\mathit{Prop}}

\renewcommand{\Diamond}{\lozenge}
\newcommand{\Diamondmulti}[1]{\Diamond^{\leq #1}}
\newcommand{\Boxmulti}[1]{\Box^{\leq #1}}

\title{The Size of Interpolants in Modal Logics}
\date{}

\author{Balder ten Cate}{University of Amsterdam}{b.d.tencate@uva.nl}{https://orcid.org/0000-0002-2538-5846}{}

\author{Louwe B. Kuijer}{University of Liverpool}{louwe.kuijer@liverpool.ac.uk}{https://orcid.org/0000-0001-6696-9023}{}

\author{Frank Wolter}{University of Liverpool}{wolter@liverpool.ac.uk}{https://orcid.org/0000-0002-4470-606X}{}

\Copyright{Balder ten Cate, Louwe B. Kuijer, and Frank Wolter}
\authorrunning{B.~ten Cate, L.~Kuijer, and F.~Wolter}

\keywords{Modal logic, strongest implicates, uniform interpolants, Craig interpolants}

\ccsdesc[500]{Theory of computation~Modal and temporal logics}
\relatedversion{A version of this paper will be presented at the 41st Annual Symposium on Logic in Computer Science (LICS 2026)}

\begin{document}

\maketitle
\begin{abstract}
   We start a systematic investigation of the 
   size of Craig interpolants, uniform interpolants, and strongest implicates for (quasi-)normal modal logics. Our main upper bound states that for tabular modal logics, the computation of strongest implicates can be reduced in polynomial time to uniform interpolant computation in classical propositional logic. Hence they are of polynomial dag-size iff NP is included in P/poly. The reduction also holds for Craig interpolants if the tabular modal logic has the Craig interpolation property. Our main lower bound shows an unconditional exponential lower bound on the size of Craig interpolants and strongest implicates covering almost all non-tabular standard normal modal logics. For normal modal logics contained in or containing S4 or GL we obtain the following dichotomy:
   tabular logics have ``propositionally sized'' interpolants while for non-tabular logics an unconditional exponential lower bound holds.  
\end{abstract}

\section{Overview}
Craig interpolation and the stronger property of uniform interpolation have been investigated extensively for modal and other logics~\cite{volume26}. The Craig interpolation property (CIP) says that whenever an implication $\varphi\rightarrow\psi$ is valid there exists a formula $\chi$ using only the shared propositional variables of $\varphi$ and $\psi$ such that both $\varphi\rightarrow\chi$ and $\chi\rightarrow \psi$ are valid. The formula $\chi$ is then called a \emph{Craig interpolant for $\varphi\rightarrow\psi$} and often provides a useful explanation of its validity~\cite{craig_1957}. Uniform interpolation (UIP) strengthens this. It says that for a signature $\sigma$ of propositional variables, a Craig interpolant $\chi$ can be chosen uniformly for all $\psi$ with $\text{sig}(\psi)\cap \text{sig}(\varphi)= \sigma$. Then $\chi$ is called a \emph{uniform $\sigma$-interpolant of $\varphi$} and can be regarded as the result of eliminating existential 
quantifiers over, or `forgetting', the variables $\text{sig}(\varphi)\setminus \sigma$ in $\varphi$~\cite{DBLP:journals/jsyml/Pitts92,visser1996uniform,DBLP:journals/apal/Liberatore24,DBLP:journals/logcom/GiessenJK25}. If a logic does not have CIP, \emph{strongest $\sigma$-implicates}, formulas $\chi$ that behave like uniform $\sigma$-interpolants but in which the condition `$\text{sig}(\psi)\cap \text{sig}(\varphi)= \sigma$' is replaced by `$\text{sig}(\psi)\subseteq \sigma$' are often a sufficient approximation of uniform interpolants. They coincide with uniform interpolants if the logic has CIP.

While existence results and effective constructions for Craig interpolants, uniform interpolants, and strongest implicates in modal logics have been studied extensively, 
size bounds often remain unknown, even for standard modal logics. Moreover, in contrast to many other properties of modal logics such as the complexity of reasoning or the size of finite models, no general results have yet been established. In this paper we address the latter problem, and show first general results on the size of interpolants. Although we are mainly interested in normal modal logics, our results can often be more conveniently formulated for the even larger class of quasi-normal modal logics, modal logics determined by classes of (possibly generalised) Kripke frames with a distinguished world. We investigate
modal logics with a single modal operator, but the extension to polymodal logics is straightforward.

The lack of general results in modal logic is not surprising: even in the case of propositional logic it is a major open problem whether Craig interpolants or uniform interpolants of polynomial size always exist. Although it is conjectured that this is false and both are of superpolynomial size in general, a proof would entail P $\not=$ NP and so is currently out of reach. As a result, the landscape of unconditional known lower bounds is rather sparse: currently, lower bounds have only been established for restricted fragments, most notably the monotone fragment of propositional logic, where one can show exponential lower bounds on the size of interpolants~\cite{razborov1985lower,DBLP:journals/combinatorica/AlonB87}.
Clearly, lower bounds obtained for propositional logic automatically transfer to modal logics. Thus, for example, the aforementioned exponential lower bound on the size of Craig interpolants in the monotone fragment of propositional logic immediately yields corresponding exponential lower bounds for monotone fragments of modal logics. An intriguing question, which we will pursue here, is when \emph{upper bounds} for the case of propositional logic transfer to the modal setting.

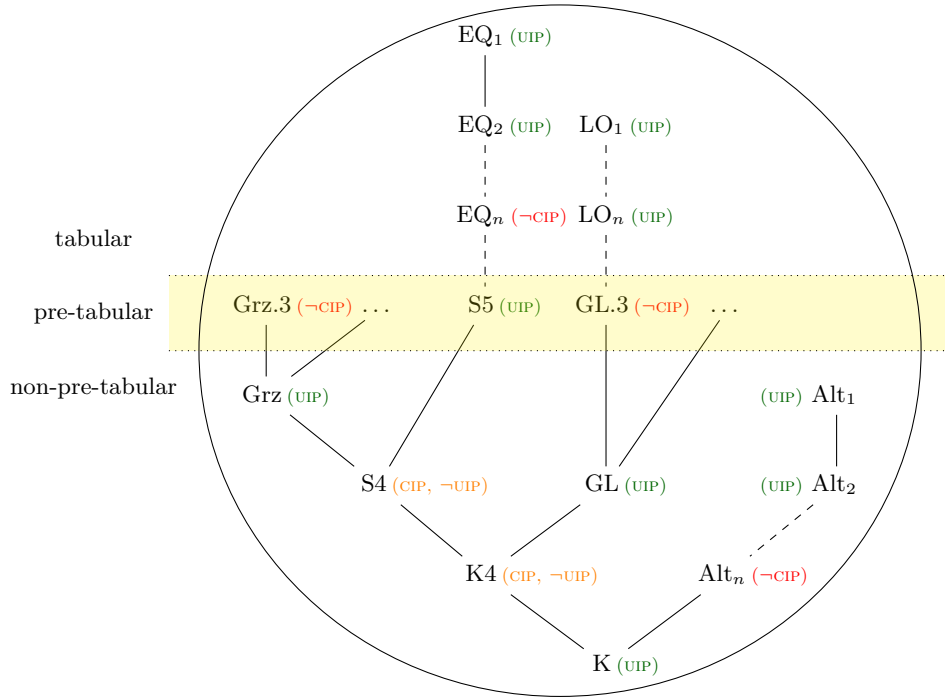
\begin{figure}[t]
\definecolor{darkgreen}{rgb}{0.0, 0.4, 0.0}
\definecolor{darkorange}{rgb}{0.9, 0.35, 0.0}
\newcommand{\noCIP}{\textsc{\color{red}\rlap{\scriptsize \!($\neg$cip)}}}
\newcommand{\hasCIP}{\textsc{\color{darkorange}\rlap{\scriptsize \!(cip, $\neg$uip)}}}
\newcommand{\hasUIP}{\textsc{\color{darkgreen}\rlap{\scriptsize \!(uip)}}}
\newcommand{\hasUIPleft}{\hspace{-7mm}\textsc{\color{darkgreen}{\scriptsize \!(uip)}}~}
\begin{center}
    \small
    \begin{tikzpicture}
    \matrix (A) [matrix of nodes, row sep=0.7cm, column sep=.6cm]
    { 
                  && EQ$_1$ \hasUIP \\
                  &            & EQ$_2$ \hasUIP & LO$_1$ \hasUIP &           \\
                  &            & EQ$_n$ \noCIP  & LO$_n$ \hasUIP &           \\
      Grz.3 \noCIP& \ldots     & S5 \hasUIP     & GL.3 \noCIP    & \ldots    \\
      Grz \hasUIP &            &                &                &  & \hasUIPleft Alt$_1$        \\
                  & S4 \hasCIP &                & GL \hasUIP     &                & \hasUIPleft Alt$_2$  \\
                  &            & K4 \hasCIP     &                & Alt$_n$ \noCIP &   \\
                  &            &                &  K \hasUIP     & \\
    };
    \draw (A-1-3)--(A-2-3);
    \draw[dashed] (A-2-3)--(A-3-3);
    \draw[dashed] (A-2-4)--(A-3-4);
    \draw[dashed] (A-3-3)--(A-4-3);
    \draw[dashed] (A-3-4)--(A-4-4);
    \draw (A-4-1)--(A-5-1);
    \draw (A-4-2)--(A-5-1);
    \draw (A-4-3)--(A-6-2);
    \draw (A-4-4)--(A-6-4);
    \draw (A-4-5)--(A-6-4);
    \draw (A-5-1)--(A-6-2);
    \draw (A-5-6)--(A-6-6);
    \draw (A-6-2)--(A-7-3);
    \draw (A-6-4)--(A-7-3);
    \draw[dashed] (A-6-6)--(A-7-5);
    \draw (A-7-3)--(A-8-4);
    \draw (A-7-5)--(A-8-4);

    \fill[fill=yellow,fill opacity=.2] (-5,0) rectangle (5.5,1);
    \draw[dotted] (-5,1) -- (5.5,1); 
    \draw[dotted] (-5,0) -- (5.5,0); 

    \node[] at (-6,-0.5) {non(-pre)-tabular};
    \node[] at (-6,0.5) {pre-tabular};
    \node[] at (-6,1.5) {tabular};

    \draw[] (0.2,0) ellipse (4.8cm and 4.6cm);

\end{tikzpicture}
\end{center}
\caption{Landscape of non-tabular, pre-tabular, and tabular normal modal logics. To keep the figure simple, not all  inclusion relations that hold between the given logics are indicated with edges. The failure of CIP for $\text{EQ}_n$ and for $\text{Alt}_n$ holds for $n>2$. 
Logics are defined formally in Section~\ref{sec:prelim} and pointers to the literature and
proofs are supplied in Section~\ref{main:notions} and Appendix~\ref{sec:smallproofs}.}
\label{fig:landscape}
\end{figure}

Not every modal logic has the CIP, let alone the UIP. Indeed, while many standard modal logics have CIP, it is, for instance, known that in the uncountable set of normal modal logics containing S4
only at most 37 normal modal logics have CIP~\cite{GabbayMaks05}. We refer the reader to Figure~\ref{fig:landscape} for a small fragment of the landscape of modal logics, containing logics without CIP, logics with CIP and logics with UIP. Here and in what follows, a logic is called tabular if it is determined by a finite set of finite frames and pre-tabular if it is not tabular but all proper extensions are tabular. Note that a tabular logic with CIP also has UIP since one can obtain a uniform $\sigma$-interpolant for a formula $\varphi$ by taking the conjunction over all relevant Craig interpolants (as there are only finitely many non-equivalent formulas over a finite signature of propositional variables).

We next discuss the main contributions of this paper. Unless stated otherwise, the size $|\varphi|$ of a formula $\varphi$ is the length of $\varphi$ if represented as a dag (or equivalently, the number of subformulas of $\varphi)$. See also Section~\ref{sec:conclusion} for further discussion regarding dag-size and tree-size.

\medskip\par\noindent\textbf{First contribution (strongest implicates and Craig interpolants for tabular logics).}
For every tabular modal logic $L$, the computation of strongest $L(\sigma)$-implicates can be reduced in poly-time to the computation of uniform interpolants in propositional logic. As a corollary, we obtain that 
    the size of strongest $L(\sigma)$-implicates is bounded polynomially in the size of propositional uniform interpolants. Let PL$(\sigma)$ and ML$(\sigma)$ denote the set of propositional and, respectively, modal formulas using variables in $\sigma$ only.
 
\begin{restatable}{theorem}{thmmain}
   \label{thm:main}
   Let $L$ be a tabular quasi-normal modal logic.
    We can associate to each signature $\sigma$ of propositional variables a companion signature $\widehat\sigma$ such that
    there are poly-time translations 
    \[
    tr_L:\text{ML}(\sigma) \to \text{PL}(\widehat\sigma) \quad \text{ and } \quad rt_L:\text{PL}(\widehat\sigma)\to \text{ML}(\sigma)
    \]
    such that for every $\phi\in \ML(\sigma)$, signature $\tau \subseteq \text{sig}(\varphi)$, and every 
    uniform $\widehat\tau$-interpolant $\psi$ of $tr_L(\phi)$ in propositional logic, $rt_L(\psi)$ is a strongest $L(\tau)$-implicate for $\phi$. 
\end{restatable}

We recall that quasi-normal modal logics form a 
generalization of normal modal logic. Hence, the above
results (and the results below) apply in particular
to normal modal logics.

\begin{corollary}
Let $L$ be a tabular quasi-normal modal logic. Let
$f:\mathbb{N}\to\mathbb{N}$ be any function such that every propositional formula $\phi$ has a uniform $\sigma$-interpolant of size at most $f(|\phi|)$ for all $\sigma\subseteq \text{sig}(\phi)$. Then every modal formula $\phi$ has a strongest $L(\sigma)$-implicate (for $\sigma\subseteq \text{sig}(\phi)$) of size at most $poly(f(poly(|\phi|)))$.
\end{corollary}

Similarly, for tabular modal logics $L$ that have CIP,  the computation of Craig interpolants in $L$ can be reduced in poly-time to the computation of Craig interpolants in propositional logic. Consequently, the size of Craig interpolants in $L$ is polynomial in the size of propositional Craig interpolants.

\begin{restatable}{theorem}{thmmainCIP}
   \label{thm:mainCIP}
    Let $L$ be a tabular quasi-normal modal logic that has CIP. 
    Given a valid modal implication $\phi\to\psi$ in $L$, and given a propositional
    Craig interpolant $\chi$ for $tr_L(\phi)\to tr_L(\psi)$, it holds that $rt_L(\chi)$ is a Craig interpolant for $\phi\to\psi$ with respect to $L$. 
\end{restatable}

Here, $tr_L$ and $rt_L$ refer to the  poly-time translation from Theorem~\ref{thm:main}.

\begin{corollary}
Let $L$ be a tabular quasi-normal modal logic that has CIP. Let
$f:\mathbb{N}\to\mathbb{N}$ be any function such that every valid propositional implication $\phi \to \psi$ has a Craig interpolant of size at most $f(|\phi\to\psi|)$. Then every modal implication $\phi\to\psi\in L$ has a Craig interpolant in $L$ of size at most $poly(f(poly(|\phi|+|\psi|)))$.
\end{corollary}

The proof uses the standard bisimulation-based characterization of Craig interpolants and strongest implicates. A key ingredient of the proof is a poly-size representation of exponentially many non-bisimilar models using \emph{abstract $\Phi$-models}, for an appropriate poly-size set $\Phi$ of propositional formulas. Intuitively, an abstract $\Phi$-model is a partial description of a Kripke model, which does not include a concrete propositional valuation for each world, but provides enough information to infer which worlds have the same propositional valuation.

\medskip\par\noindent\textbf{Second contribution (exponential lower bounds for non-tabular logics)}
 We prove a general exponential lower bound on the size of Craig interpolants and strongest implicates, which applies to many logics not covered by Theorem~\ref{thm:main}. 
 
\begin{restatable}{theorem}{thmmainlower}
\label{thm:main-lower}
 Let $L$ be any non-tabular normal modal logic that contains or is contained in S4 or the G\"{o}del-L\"{o}b provability logic GL, or, more generally, has a non-tabular intersection with S4 or GL.
 \begin{enumerate}
     \item
 There 
 exist modal formulas $(\phi_n)_{n=1,2,\ldots}$ of size polynomial in $n$ 
 and signatures $(\sigma_n)_{n=1,2,\ldots}$
 with $\sigma_n\subseteq sig (\phi_n)$, such that $\phi_n$ has a strongest $L(\sigma_n)$-implicate, and
  every strongest $L(\sigma_n)$-implicate of $\phi_n$ has size at least $2^n$.
    \item There are modal formulas $(\phi_n)_{n=1,2,\ldots}$
 and 
 $(\psi_n)_{n=1,2,\ldots}$
 of size polynomial in $n$ 
 such that $\phi_{n}\to\psi_{n}\in L$ for all $n\geq 1$, a Craig interpolant for $\phi_{n} \to \psi_{n}$ exists for all $n\geq 1$, and every Craig interpolant for $\phi_{n}\to\psi_{n}$ in $L$ 
 has size at least $2^n$.
 \end{enumerate}
 
\end{restatable}
The proof is based on a complete description of the pre-tabular normal modal logics containing S4 or GL and exploits their \emph{poly-size model property}. In the case of strongest implicates, we actually prove a more general sufficient condition for an exponential lower bound that introduces the following \emph{exponential growth property} of a modal logic:
there are frames validating the logic in which one can reach exponentially many worlds in a polynomial number of steps from a distinguished world. All pre-tabular normal modal logics containing S4 or GL enjoy that property. 

Interestingly, Theorem~\ref{thm:main} and~\ref{thm:mainCIP} and 
Theorem~\ref{thm:main-lower} together provide a dichotomy for normal modal  logics $L$ that are contained in or contain S4 or GL:
\begin{itemize}
    \item If $L$ is tabular, then $L$ admits ``propositional sized'' strongest $L(\sigma)$-implicates and (if $L$ has CIP) Craig interpolants;
    \item If $L$ is non-tabular, there are exponential lower bounds on the size of strongest $L(\sigma)$-implicates and Craig interpolants. 
\end{itemize}
Note that tabularity is decidable for normal modal logics containing S4 or GL, so this dichotomy is effective~\cite[Chapter~17]{DBLP:books/daglib/0030819}.

We note that from Theorem~\ref{thm:main-lower} we obtain tight bounds for the size of Craig interpolants for many standard modal logics, including K, K4, S4, GL, Grzegorczyk's logic Grz, and S5. 
\begin{theorem}
    Let $L$ be either K, K4, S4, Grz, GL, or S5. Then for every $\varphi \rightarrow \psi\in L$ there exists a Craig interpolant of at most exponential size. This bound is optimal. 
\end{theorem}
The exponential upper bounds for K and S5 are shown in~\cite{TenEtAl06,GhilardiLWZ06} (even for uniform interpolants) and they follow from~\cite{krachtbook,TenEtAl13} for the remaining logics. The exponential lower bound for GL was independently shown in~\cite{DBLP:conf/csl/PapafilippouF25}, where exponential lower bounds for K4 and S4 are mentioned as open research problems.  

Finally, we refute the natural conjecture that the above dichotomy for S4 and GL generalizes to all normal modal logics. We show that the non-tabular logic $\text{Alt}_1$ determined by frames in which every
world has at most one successor admits a poly-time reduction of Craig and uniform interpolants to propositional Craig and, respectively, uniform interpolants.

The paper is structured as follows. In Section~\ref{sec:prelim} we introduce the relevant notation and results in modal logic. Next, in Section~\ref{main:notions}, we discuss the main notions investigated in this paper and also review related work in propositional and modal logic.
Section~\ref{sec:toprop} is the core of this paper where we present the reduction of strongest implicates in tabular modal logics to uniform interpolants in propositional logic. The results and techniques presented in Section~\ref{sec:toprop} are then also applied to give the reduction for Craig interpolants in tabular modal logics with CIP to propositional Craig interpolants in Section~\ref{sec:CraigInt}. Exponential lower bounds for interpolants are presented in Section~\ref{sec:lower}. Details of some proofs are given in the appendix. 
\section{Preliminaries}
\label{sec:prelim}
We start by introducing relevant notions from modal and propositional logic~\cite{DBLP:books/daglib/0030819,Blackburn_Rijke_Venema_2001}.
Fix a countable infinite set $\Prop$ of \emph{atoms} or \emph{propositional variables}. 
A \emph{signature} $\sigma$ is a finite subset of $\Prop$.
The language PL of \emph{propositional formulas} consists of all formulas constructed from atoms using 
the propositional connectives $\top$, $\neg$, and $\wedge$ in the usual way.
Similarly, the language ML of \emph{modal formulas} consists of formulas constructed from atoms using the propositional connectives $\top$, $\neg$, and $\wedge$
as well as the  modal operator $\Box$. We set $\Diamond\varphi=\neg\Box\neg\varphi$, define $\Box^{\leq n}\varphi=\varphi \wedge \Box\varphi \wedge \cdots \wedge \Box^{n}\varphi$, and let $\Diamond^{\leq n}\varphi=\neg \Box^{\leq n}\neg \varphi$,
for $n\geq 0$.
By $\text{sig}(\varphi)$ we denote the set of atoms in the (propositional or modal) formula $\varphi$.
For a given signature $\sigma$, 
PL$(\sigma)$ consists of those formulas 
$\varphi\in \PL$ with $\text{sig}(\varphi)\subseteq\sigma$,
and ML$(\sigma)$ again consists of those formulas 
$\varphi\in \ML$ with $\text{sig}(\varphi)\subseteq\sigma$.  

Models for propositional formulas are defined as usual. For a propositional formula $\varphi$, we write $\models \varphi$ if $\varphi$ is valid.
ML$(\sigma)$ is interpreted in \emph{frames} $\F=(W,R)$ with $W$ a nonempty set of worlds and $R\subseteq W \times W$ an accessibility relation. A \emph{pointed frame} is a pair $(\F,w)$ with $w\in W$. A world $w\in W$ is called a \emph{root} of $\F$ if for every $v\in W$ there exists an $R$-path from $w$ to $v$. A pointed frame $(\F,w)$ with $w$ a root of $\F$ is called \emph{rooted}. The \emph{size} $|\F|$ of a frame $\F$ is the number of worlds in $\F$. 

A \emph{model} based on $\F$ is a pair $\M=(\F,V)$ with $V:W\rightarrow 2^\Prop$ a \emph{valuation}. The \emph{satisfaction-relation} $\models$ between \emph{pointed models} $(\M,w)$ with $w\in W$
and modal formulas $\varphi$ is defined as follows:
\begin{itemize}
    \item $\M,w\models p$ if $p\in V(w)$;
    \item $\M,w\models \neg \varphi$ if $\M,w\not\models\varphi$;
    \item $\M,w\models \varphi \wedge \psi$ if $\M,w\models \varphi$ and $\M,w\models \psi$;
    \item $\M,w\models \Box \varphi$ if $\M,v\models \varphi$ for all $v$ with $wRv$.
\end{itemize}
We write $\F,w\models \varphi$ and call $\varphi$ \emph{valid in $(\F,w)$} if $\M,w\models \varphi$ for all models $\M$ based on $\F$. For $w\in W$ denote by $\F_{w}$ the restriction of $\F$ to the set of worlds $v\in W$ such that there is an $R$-path from $w$ to $v$. Then $w$ is a root of $\F_{w}$ and one can easily show that $\F,w\models \varphi$ iff $\F_{w},w\models \varphi$. 

Let $\mathcal{F}$ be a set of pointed frames. Then 
\[
\text{Log}(\mathcal{F})=\{\varphi\in \text{ML} \mid \text{ for all $(\F,w)\in \mathcal{F}: \F,w\models \varphi$}\}
\]
is the \emph{logic determined by $\mathcal{F}$}. Note that Log$(\mathcal{F})$ is always a \emph{quasi-normal modal logic}, a very general class of modal logics defined as follows~\cite{DBLP:books/daglib/0030819}. The smallest quasi-normal modal logic is defined as $\text{K} =\text{Log}(\mathcal{F}_{K})$, where $\mathcal{F}_{K}$ is the class of all pointed frames. Then a quasi-normal modal logic $L$ is any set of modal formulas containing K and closed under modus ponens and uniform substitutions of atoms by modal formulas.
We note that there are quasi-normal modal logics that are not of the form Log$(\mathcal{F})$ for any class $\mathcal{F}$ of pointed frames, but they will not play any important role in this paper.
Observe that Log$(\mathcal{F})=\text{Log}(\{(\F_{w},w)\mid (\F,w)\in \mathcal{F}\})$, so we can always work with rooted frames instead of arbitary pointed frames. If $L=\text{Log}(\mathcal{F})$ for some class $\mathcal{F}$ such that 
for every $(\F,w)\in \mathcal{F}$ the set $\mathcal{F}$ also contains all $(\F_{v},v)$ with $v$ in $\F$, then Log$(\mathcal{F})$ is a \emph{normal modal logic}, a quasi-normal modal logic that is also closed under the rule: if $\varphi\in L$, then $\Box\varphi\in L$. For normal modal logics $L$ we often drop the distinguished worlds when defining $L$ as Log$(\mathcal{F})$. Table~\ref{tab:logics} defines relevant normal modal logics. 

\begin{table*}
\begin{center}
\begin{tabular}{|l | l|} 
 \hline
 Log$(\mathcal{F})$ & Class $\mathcal{F}$ of frames $\F=(W,R)$ \\ [0.5ex] 
 \hline
 K & all frames \\ 
 Alt$_n$ & frames in which each node has at most $n$ successors \\ 
 S4 & transitive and reflexive frames \\
 Grz & frames validating S4 and no infinite strict chains $w_{0}Rw_1 R w_{2} \cdots$ \\
 GL & transitive and no infinite chains $w_{0}R w_{1} R w_{2} \cdots$ \\
 S5 & equivalence relations \\ 
 Grz.3/GL.3 & GL/Grz frames satisfying $\forall w\forall v (wRv \vee vRw \vee w=v)$ \\
 LO$_{n}$  & GL.3 frames of length at most $n$ \\
 EQ$_{n}$   &  S5 frames of size at most $n$ \\
 [1ex]
 \hline
\end{tabular}
\end{center}
\caption{List of relevant normal modal logics}
\label{tab:logics}
\end{table*}

We refer the reader to~\cite{DBLP:books/daglib/0030819} for a detailed discussion of \mbox{(quasi-)} normal modal logics.
A quasi-normal modal logic $L$ is \emph{tabular} if there exists a finite set $\mathcal{F}$ of finite pointed frames such that $L=\text{Log}(\mathcal{F})$.

Let $\Phi$ be a set of formulas and let $\M_{i}=(\F_{i},V_{i})$ be models with
$\F_{i}=(W_{i},R_{i})$, for $i=1,2$. Then $\M_{1},w_{1}$ and $\M_{2},w_{2}$ are \emph{$\Phi$-indistinguishable}, in symbols $\M_{1},w_{1} \equiv_{\Phi} \M_{2},w_{2}$, if $\M_{1},w_{1}$
and $\M_{2},w_{2}$ satisfy the same formulas in $\Phi$.
Let $\sigma$ be a signature. Then ML$(\sigma)$-indistinguishability is characterized using $\sigma$-bisimilations. A \emph{$\sigma$-bisimulation} between $\M_{1}$ and $\M_{2}$ is any
relation $Z\subseteq W_{1} \times W_{2}$ such that for all $w_{1}\in W_{1}$ and $w_{2}\in W_{2}$:
\begin{description}
    \item[Atoms] if $w_{1} Z w_{2}$, then $\M_{1},w_{1}\models p$ iff $\M_{2},w_{2}\models p$ for all $p\in \sigma$,
    \item[Forth] if $w_{1} Z w_{2}$ and $(w_{1},v_{1})\in R_{1}$, then there is a $v_{2}\in W_{2}$ such that $(w_{2},v_{2})\in R_{2}$ and $v_{1} Z v_{2}$,
    \item[Back] if $w_{1} Z w_{2}$ and $(w_{2},v_{2})\in R_{2}$, then there is a $v_{1}\in W_{1}$ such that $(w_{1},v_{1})\in R_{1}$ and $v_{1} Z v_{2}$.
\end{description}
If $Z$ is a $\sigma$-bisimulation between $\M_1$ and $\M_2$ such that $w_1Zw_2$, we write $Z: \M_1,w_1\bisim_\sigma \M_2,w_2$. If there is any $Z$ such that $Z: \M_1,w_1\bisim_\sigma \M_2,w_2$ we write $\M_1,w_1\bisim_\sigma \M_2,w_2$ and call $\M_{1},w_{1}$ and $\M_{2},w_{2}$ \emph{$\sigma$-bisimilar}. It is known that $\bisim_{\sigma}$ and $\equiv_{\ML(\sigma)}$ coincide in finite pointed models~\cite{Blackburn_Rijke_Venema_2001}.

\section{Main Notions}
\label{main:notions}
In this paper, our main concern is the size and computation of strongest $\sigma$-implicates, uniform interpolants, and Craig interpolants. We next introduce these notions in detail in both modal and propositional logic.

Consider a quasi-normal logic $L$, a modal formula $\varphi$ and a signature $\sigma\subseteq \text{sig}(\varphi)$. A modal formula $\chi$ is called a \emph{strongest $L(\sigma)$-implicate of $\varphi$} if $\text{sig}(\chi)=\sigma$ and
\begin{itemize}
    \item $\varphi\rightarrow \chi\in L$;
    \item if $\varphi \rightarrow \psi \in L$ and $\text{sig}(\psi)\subseteq \sigma$, then
    $\chi \rightarrow \psi\in L$.
\end{itemize}
By definition, any two strongest $L(\sigma)$-implicates of $\varphi$ are logically equivalent in $L$, so we speak about \emph{the} strongest $L(\sigma)$-implicate if it exists. In this paper, we are interested in the computation and size of strongest $L(\sigma)$-implicates. There is a close relationship between strongest implicates and uniform interpolants. Recall that a formula $\chi$ is called a \emph{Craig interpolant in $L$} for an implication $\varphi \rightarrow \psi\in L$ if $\varphi \rightarrow \chi\in L$,
$\chi \rightarrow \psi\in L$, and $\text{sig}(\chi) \subseteq \text{sig}(\varphi) \cap \text{sig}(\psi)$. A logic $L$ has the \emph{Craig interpolation property (CIP)} if for any $\varphi \rightarrow \psi\in L$ a Craig interpolant exists. A formula $\chi$ is called a \emph{uniform $L(\sigma)$-interpolant} for $\varphi$ if it satisfies the conditions of a strongest $L(\sigma)$-implicate of $\varphi$ even if Condition 2 is strengthened by replacing the condition `$\text{sig}(\psi)\subseteq \sigma$' by the condition $\text{sig}(\psi) \cap \text{sig}(\varphi)\subseteq \sigma$. It follows that a uniform $L(\sigma)$-interpolant for $\varphi$ is a Craig interpolant for all $\varphi \rightarrow \psi\in L$ with $\text{sig}(\psi) \cap \text{sig}(\varphi)\subseteq \sigma$. The following is straightforward. 
\begin{lemma}
 If $L$ has CIP, then strongest $L(\sigma)$-implicates and uniform $L(\sigma)$-interpolants coincide: if $\chi$ is a strongest $L(\sigma)$-implicate of $\varphi$, then it is a uniform $L(\sigma)$-interpolant for $\varphi$.
\end{lemma}

\begin{example}\label{example:first}
Let $L_{\G}=\text{Log}(\{(\G,0)\})$ with $\G$ the frame below:
\begin{center}
\begin{tikzpicture}[xscale=1, yscale=0.7]
    \node[draw,circle, inner sep = 2pt] (s0) at (0,0) {$0$};
    \node[draw,circle, inner sep = 2pt] (s1) at (1.5,1) {$1$};
    \node[draw,circle, inner sep = 2pt] (s2) at (1.5,0) {$2$};
    \node[draw,circle, inner sep = 2pt] (s3) at (1.5,-1) {$3$};
	
	\draw[->] (s0) -- (s1);
	\draw[->] (s0) -- (s2);
    \draw[->] (s0) -- (s3);
\end{tikzpicture}
\end{center}
The logic $L_{\G}$ does not have the CIP. For example, the strongest $L_{\G}(\{p\})$-implicate of $\lozenge (p\wedge q)\wedge \lozenge (p\wedge \neg q)$ is $\lozenge p$. Yet $\lozenge p$ is not a uniform $L_{\G}(\{p\})$-interpolant, since $(\lozenge (p\wedge q)\wedge \lozenge (p\wedge \neg q))\rightarrow (\lozenge (\neg p\wedge r)\rightarrow \square (\neg p\rightarrow r))\in L_{\G}$ while $\lozenge p \rightarrow (\lozenge (\neg p\wedge r)\rightarrow \square (\neg p\rightarrow r))\not\in L_{\G}$.
\end{example}
The following result characterizes strongest $L(\sigma)$-implicates using bisimulations: a
strongest $L(\sigma)$-implicate of $\varphi$ expresses existentially quantifying $\text{sig}(\varphi)\setminus \sigma$ \emph{modulo} a $\sigma$-bisimulation. Similar characterisations have been shown for many modal logics~\cite{DBLP:journals/jsyml/Pitts92,visser1996uniform,MU-CALCULUS,FOUNDATIONS_ALC_UI}.

\begin{theorem}\label{thm:bisim}
    Let $L=\text{Log}(\mathcal{F}_{L})$ with $\mathcal{F}_{L}$ a finite sets of finite pointed frames. Let $\varphi$ be a modal formula and $\sigma\subseteq \text{sig}(\varphi)$. Then the following conditions are equivalent for all modal formulas $\chi$ with $\text{sig}(\chi)\subseteq \sigma$: 
    \begin{enumerate}
        \item $\chi$ is the strongest $L(\sigma)$-implicate of $\varphi$;
        \item for all models $(\M,w)$ based on some $(\F,w)\in \mathcal{F}_{L}$: $\M,w\models \chi$ iff there exists a model $(\M',w')$ based on some $(\F',w')\in \mathcal{F}_{L}$ with $\M',w'\bisim_{\sigma} \M,w$ and $\M',w'\models \varphi$.
    \end{enumerate}  
\end{theorem}
The definitions of strongest $\sigma$-implicates, uniform interpolants, and Craig interpolants in propositional logic are derived from the definitions for modal logic in the obvious way. As propositional logic has CIP,
strongest propositional $\sigma$-implicates are uniform propositional $\sigma$-interpolants. 

Define the \emph{dag-size} $|\varphi|$ of a propositional or modal formula $\varphi$ as the
number of its distinct subformulas. This corresponds to representing formulas as directed acyclic graphs and is the standard measure used in circuit complexity theory. If we take into account the number of different occurrences of subformulas, we obtain the \emph{tree-size} $s(\varphi)$ of $\varphi$. Clearly, $s(\varphi)\geq |\varphi|$ and it is conjectured that in propositional logic there is a superpolynomial (even exponential) gap between 
dag-size and tree-size. However, this conjecture remains open~\cite{DBLP:journals/siamcomp/Rossman18}.

It is also conjectured that both propositional Craig interpolants and uniform interpolants are (in the worst case) necessarily of exponential dag-size. In more detail, the following result shows how the size of propositional interpolants is related to open problems in circuit
complexity~\cite{koopmann2025interpolationclassicalpropositionallogic}, see also \cite{DBLP:journals/aml/Mundici83,DBLP:conf/dagstuhl/SchoningT06}.
\begin{theorem}~\label{thm:uniformequiv}

(1) Propositional logic has poly-dag-size uniform interpolants if, and only if, $\text{NP}\subseteq \text{P}/\text{poly}$.

(2) If propositional logic has poly-dag-size Craig interpolants, then $\text{NP}\cap \text{coNP}\subseteq \text{P}/\text{poly}$.
\end{theorem}
We remind the reader that the question whether 
$\text{NP}\subseteq \text{P}/\text{poly}$ is a long-standing open problem. It is unlikely that it holds since, by the Karp-Lipton Theorem, if $\text{NP}\subseteq \text{P}/\text{poly}$, then the polynomial hierarchy collapses at the second level~\cite{karp1980some,karp1982turing,DBLP:books/daglib/0023084}.
Conversely, if we could prove $\text{NP}\not\subseteq \text{P}/\text{poly}$, then
$\text{NP} \not= \text{P}$ would follow since $\text{P}\subseteq \text{P}/\text{poly}$.
It is also regarded as unlikely that $\text{NP}\cap \text{coNP}\subseteq \text{P}/\text{poly}$. However, if we could prove $\text{NP}\cap\text{coNP}\not\subseteq \text{P}/\text{poly}$, then again $\text{NP} \not= \text{P}$ would follow. 

We next review relevant results on interpolants in modal logic.
Among the main modal logics considered in this paper, K, S5, GL, and Grz have UIP~\cite{visser1996uniform} while S4 and K4 do not~\cite{DBLP:journals/sLogica/GhilardiZ95,DBLP:journals/sLogica/Bilkova07}. S4 and K4 have CIP, however, while GL.3 and Grz.3 do not even enjoy CIP~\cite{GabbayMaks05}. Regarding the size of Craig/uniform interpolants for the logics with CIP and UIP considered in this paper, for modal logics K and S5 it is known that uniform interpolants of at most exponential size always exist~\cite{TenEtAl06,GhilardiLWZ06}. In contrast, for GL and Grz, to date, only non-elementary methods are known for constructing uniform interpolants. Interestingly, however, there are tools available for computing uniform interpolants in these logics~\cite{DBLP:conf/ijcar/FereeGGS24}. For Craig interpolants, for K and S5, the exponential upper bound is inherited from the exponential upper bound for uniform interpolants. For K4, S4, Grz, and GL one can obtain exponential upper bounds by using the reductions to Craig interpolants in K given in~\cite[Section~3.7]{krachtbook} and applying the tableau 
developed in~\cite{TenEtAl13}. While this is done explicitly only for K4, it is straightforward to extend the proof to S4, Grz, and GL. 

It is also worth noting that it is undecidable whether a finitely axiomatized normal modal logic has the CIP~\cite[Section~9.6]{krachtbook}. By the same technique, one can show that the UIP is likewise undecidable. However, this does not preclude the possibility of establishing effective dichotomies for the size of interpolants. Indeed, even though CIP is undecidable for finitely axiomatized normal modal logics containing GL \cite[Theorem~17.19]{DBLP:books/daglib/0030819}, we are nonetheless able to prove a (conditional) dichotomy result for interpolant size for normal modal logics containing GL. This dichotomy is effective since it is determined by tabularity which is effective for finitely axiomatized normal modal logics containing GL. 

\section{Constructing Strongest Implicates via Propositional Logic}
\label{sec:toprop}

We show how one can construct, for tabular quasi-normal logics, in polynomial time 
strongest implicates from uniform interpolants in propositional logic. 
\subsection{Abstraction}
The number of different valuations $V$ of atoms in a signature $\sigma$ in a fixed finite
frame is exponential in $|\sigma|$, even for frames with a single world. To avoid working with all valuations, in this subsection we introduce abstract models, of which we only require polynomially many to supply sufficient information about the underpinning models for our purposes. Unless stated otherwise, in this section, we only consider finite rooted pointed frames $(\F,w)$. We typically denote the root $w$ by $0$.

Given a signature $\sigma$,
we call a finite set $\Phi\subseteq \text{PL}(\sigma)$ an \emph{exhaustive and mutually exclusive set over $\sigma$}, $\sigma$-EME for short, if (i) $\models \bigvee_{\phi\in \Phi}\phi$ and (ii) $\models \neg (\phi\wedge \psi)$ for all $\phi\not=\psi$ with $\phi,\psi\in \Phi$. If $\M=(\F,V)$ is a model and $\Phi$ a $\sigma$-EME, we say that $\Phi$ is a \emph{$\sigma$-cover} of $\M$ if for all $\phi\in \Phi$ and all $w_1,w_2\in W$: if $\M,w_1\models \phi$ and $\M,w_2\models \phi$ then $\M,w_1\equiv_{\text{PL}(\sigma)} \M,w_2$.

If $\Phi$ is a $\sigma$-EME, then every world of $\M$ satisfies exactly one formula $\phi\in \Phi$. So $\Phi$ is a $\sigma$-cover of $\M$ if, and only if, any two worlds of $\M$ that are distinguishable in $\text{PL}(\sigma)$ are distinguished by $\Phi$.

\begin{lemma}
\label{lemma:inv}
For every $\sigma$-EME $\Phi$ and
for every 
rooted model $\M,0$ of size at most $N$,
we have  $\Phi$ is a $\sigma$-cover of $\M$ if, and only if, 
$\M,0\models \inv(\sigma,\Phi)$ 
where
\[\inv(\sigma,\Phi) = {} 
\bigwedge_{\phi\in \Phi}\bigwedge_{p\in \sigma}\left(\Diamondmulti{N}(\phi\wedge p)\rightarrow\Boxmulti{N}(\phi\rightarrow p)\right)
\]
\end{lemma}
The proof is straightforward, so we omit it here.

A key observation that we will make use of is
that, even though the number of possible $\sigma$-valuations over a given frame is exponential in $|\sigma|$, the number of $\sigma$-EMEs needed in order to cover all of them is only polynomial in $|\sigma|$.

\begin{restatable}{lemma}{lemmacoveramount}
\label{lemma:cover_amount}
Fix a finite frame $\F=(W,R)$. For every signature $\sigma$, we can compute in polynomial time a set $\Gamma_\F$ of $\sigma$-EMEs such that every model based on $\F$ has a $\sigma$-cover in $\Gamma_\F$. Moreover, each $\Phi\in\Gamma_\F$ consists of at most 
$|W|$ formulas, each of
size $O(|W|)$.
\end{restatable}

The proof (given in the appendix) is along the following lines: in any model based on $\F$, there are at most $|W|$ distinguishable worlds. Using a partition refinement procedure we can find, for each indistinguishability class, an identifying formula that is a conjunction of literals over $\sigma$ of size at most $O(|W|)$. Hence every model has a $\sigma$-cover $\Phi$ such that $|\Phi|\leq |W|$ and $|\phi|\leq O(|W|)$ for each $\phi\in \Phi$. 
Because the bounds on $|\Phi|$ and $|\phi|$ do not depend on $\sigma$, each such cover contains only a constant number of atoms. There are polynomially many ways to choose a constant number of atoms from $\sigma$, so $\Gamma_\F$ can be computed in polynomial time (and $|\Gamma_\F|$ is polynomial).

\begin{example} Assume $\sigma=\{p,q\}$. If $\F$ has a single world, then we can take $\Gamma_{\F}=\{\{\top\}\}$ and if $\F$ has two worlds, $\Gamma_{\F}=\{\{p,\neg p\},\allowbreak \{q,\neg q\},\allowbreak \{\top\}\}$ is as required.
\end{example}
If $\Phi$ is a $\sigma$-EME, then an \emph{abstract $\Phi$-model} is a pair $\A=(\F,f)$, where $\F=(W,R)$ is a frame and $f:W\rightarrow \Phi$ is a function. One can regard abstract $\Phi$-models as partially specified models: rather than stating for every world $w$ and atom $p$ whether $p$ is true at $w$, they only state which element of $\Phi$ is true in $w$. 
If $\Phi$ is a $\sigma$-cover of $\M$, then the \emph{$\Phi$-abstraction of $\M$} is the unique abstract $\Phi$-model $\A=(\F,f)$ such that $\M,w\models f(w)$ for all $w\in W$. 

We also lift bisimulations from the level of models to the level of abstract $\Phi$-models.
Let $\Phi$ be a $\sigma$-EME and $\A_{i}=(\F_{i},f_{i})$ be abstract $\Phi$-models, for $i=1,2$. A relation $Z \subseteq W_{1}\times W_{2}$ is an \emph{abstract $\Phi$-bisimulation} between $\A_{1}$ and $\A_{2}$ 
if conditions {\bf Forth} and {\bf Back} introduced in the definition of $\sigma$-bisimulations hold and 

\begin{description}
    \item[Abstract atoms] for all $w_{1}\in W_{1}$ and $w_{2}\in W_{2}$, if $w_{1}Z w_{2}$, then $f_{1}(w_{1})=f_{2}(w_{2})$.
\end{description}

We write $Z: \A_1,w_1\bisim_\Phi \A_2,w_2$ if $Z$ is an abstract $\Phi$-bisimulation such that $w_1Zw_2$. If there is a $Z$ such that $Z: \A_1,w_1\bisim_\Phi \A_2,w_2$ we write
$\A_{1},w_{1}\bisim_{\Phi}\A_{2},w_{2}$ and say that $\A_{1},w_{1}$ and $\A_{2},w_{2}$ are \emph{abstractly $\Phi$-bisimilar}.

\begin{example}\label{ex:abstract}Consider the following two models $\M_1$ and $\M_2$, based on the frame $\G$ from Example~\ref{example:first}.
\begin{center}
\begin{tikzpicture}
    \node[draw,circle, inner sep = 2pt, label=left:$r$] (s0) at (0,0) {$0$};
    \node[draw,circle, inner sep = 2pt, label=right:{$p, q$}] (s1) at (1.5,1) {$1$};
    \node[draw,circle, inner sep = 2pt, label=right:{$p$}] (s2) at (1.5,0) {$2$};
    \node[draw,circle, inner sep = 2pt, label=right:$s$] (s3) at (1.5,-1) {$3$};
	
	\draw[->] (s0) -- (s1);
	\draw[->] (s0) -- (s2);
    \draw[->] (s0) -- (s3);
\end{tikzpicture}\hspace{5pt}
\begin{tikzpicture}
    \node[draw,circle, inner sep = 2pt, label=left:$r$] (s0) at (0,0) {$0$};
    \node[draw,circle, inner sep = 2pt, label=right:{}] (s1) at (1.5,1) {$1$};
    \node[draw,circle, inner sep = 2pt, label=right:{}] (s2) at (1.5,0) {$2$};
    \node[draw,circle, inner sep = 2pt, label=right:{$p$}] (s3) at (1.5,-1) {$3$};
	
	\draw[->] (s0) -- (s1);
	\draw[->] (s0) -- (s2);
    \draw[->] (s0) -- (s3);
\end{tikzpicture}
\end{center}
Let $\sigma = \{p,r,s\}$, $\phi_1  = p$, $\phi_2=\neg p \wedge r$ and $\phi_3=\neg p \wedge \neg r$. Then $\Phi = \{\phi_1,\phi_2,\phi_3\}$ is a $\sigma$-cover of both $\M_1$ and $\M_2$. Furthermore, their $\Phi$-abstractions, $\A_1$ and $\A_2$, are shown below. 
\begin{center}
\begin{tikzpicture}
    \node[draw,circle, inner sep = 2pt, label=left:$\phi_2$] (s0) at (0,0) {$0$};
    \node[draw,circle, inner sep = 2pt, label=right:{$\phi_1$}] (s1) at (1.5,1) {$1$};
    \node[draw,circle, inner sep = 2pt, label=right:$\phi_1$] (s2) at (1.5,0) {$2$};
    \node[draw,circle, inner sep = 2pt, label=right:$\phi_3$] (s3) at (1.5,-1) {$3$};
	
	\draw[->] (s0) -- (s1);
	\draw[->] (s0) -- (s2);
    \draw[->] (s0) -- (s3);
\end{tikzpicture}\hspace{15pt}
\begin{tikzpicture}
    \node[draw,circle, inner sep = 2pt, label=left:$\phi_2$] (s0) at (0,0) {$0$};
    \node[draw,circle, inner sep = 2pt, label=right:{$\phi_3$}] (s1) at (1.5,1) {$1$};
    \node[draw,circle, inner sep = 2pt, label=right:$\phi_3$] (s2) at (1.5,0) {$2$};
    \node[draw,circle, inner sep = 2pt, label=right:$\phi_1$] (s3) at (1.5,-1) {$3$};
	
	\draw[->] (s0) -- (s1);
	\draw[->] (s0) -- (s2);
    \draw[->] (s0) -- (s3);
\end{tikzpicture}
\end{center}
The models $\M_1$ and $\M_2$ are not $\sigma$-bisimilar, but their abstractions $\A_1$ and $\A_2$ are abstractly $\Phi$-bisimilar, as witnessed by $Z=\{(0,0), (1,3), (2,3), (3,1), (3,2)\}$.
\end{example}
While $Z$ is not a $\sigma$-bisimulation between $\M_1$ and $\M_2$, we will later rely on the fact that there is a different model $\M_2'$ such that $Z$ is a $\sigma$-bisimulation between $\M_1$ and $\M_2'$.

\begin{restatable}{proposition}{propsinglemodelbisim}
\label{prop:single_model_bisim}
Let $\M,0$ be a rooted model, $\Phi$ a $\sigma$-cover of $\M$ and $\A$ the $\Phi$-abstraction of $\M$. Then for every rooted frame $(\F',0)$ with $\F'=(W',R')$ and for every $Z\subseteq W\times W'$ the following are equivalent:
\begin{enumerate}
	\item $Z:(\M,0)\bisim_{\sigma} (\M',0)$ for some model $\M'$ based on $\F'$, 
	\item $Z: (\A,0)\bisim_\Phi (\A',0)$ for some abstract $\Phi$-model $\A'$ based on $\F'$.
\end{enumerate}
Furthermore, if such $\A'$ and $\M'$ exist then $\A'$ is the $\Phi$-abstraction of $\M'$. 
\end{restatable}
A detailed proof can be found in the appendix.

For a $\sigma$-EME $\Phi$, we often treat the propositional formulas in $\Phi$ as if they are atoms. Specifically,
we will denote by ML$(\Phi)$
the set of modal formulas constructed from the formulas in $\Phi$ (treated as atoms) in the usual way. 
In abstract $\Phi$-models $\A$, formulas $\psi\in \text{ML}(\Phi)$ are evaluated in the expected way by setting $\A,w\models \phi$ iff $f(w)=\phi$ for all $\phi\in \Phi$ and then using the standard truth 
conditions for modal formulas.
Observe that if $\psi\in \text{ML}(\Phi)$ and $\A$ is the $\Phi$-abstraction of $\M$, then $\M,w\models \psi$ if, and only if, $\A,w\models \psi$.

\begin{restatable}{proposition}{propidentifiersize}
\label{prop:identifier_size}
Fix a natural number $N>0$.
Given a $\sigma$-EME $\Phi$ (with $\sigma$ a signature) and a rooted abstract $\Phi$-model $(\A,0)$ of size at most $N$,
we can compute in polynomial time
a formula $\delta_\A\in \text{ML}(\Phi)$ such that,
for all rooted abstract $\Phi$-models $(\A',0)$ of size at most $N$,
$\A',0\models\delta_\A$ iff
$\A',0\bisim_{\Phi} \A,0$.
Moreover, $|\delta_\A|\leq O(2^N\cdot\max_{\phi\in\Phi}|\phi|)$. 
\end{restatable}

The proof (given in the appendix) is along the following lines: let $\A=(\F,f)$ be given. For each $w\in W$, let $[w] = \{w'\mid \A,w\bisim_\Phi \A,w'\}$. We first use a partition refinement procedure to compute, in polynomial time, a formula $\psi_{[w]}\in \text{ML}(\Phi)$ such that $\A,w'\models \psi_{[w]}$ iff $w'\in [w]$.

This procedure works in the same way one would construct a characteristic formula for a $\sigma$-bisimulation class of a finite model. That is to say, we start with the labeled partition
\(\mathcal{I}_0=\{(W_\psi,\psi)\mid \psi\in \Phi\},\)
where $W_\psi=\{w\in W\mid \A,w\models \psi\}$. Now, consider any $(J,\psi), (J',\psi')\in \mathcal{I}_m$, and let $J^+=\{j\in J\mid \exists j'\in J': (j,j')\in R\}$ and $J^-=\{j\in J\mid \forall j'\in J: (j,j')\not\in R\}$. If both $J^+$ and $J^-$ are non-empty, we obtain $\mathcal{I}_{m+1}$ from $\mathcal{I}_m$ by replacing $(J,\psi)$ with $(J^+,\psi\wedge \lozenge \psi')$ and $(J^-,\psi\wedge\neg\lozenge \psi')$.

Once no further refinements are possible, which happens after at most $N$ steps, the labeled partition will be of the form $\mathcal{I}=\{([w],\psi_{[w]})\mid w\in W\}$, where $\A,w'\models \psi_{[w]}$ if and only if $w'\in [w]$.
We then define
\[\delta_\A = \psi_{[0]}\wedge \Big( \Boxmulti{N} \bigvee_{w\in W}\psi_{[w]} \Big) \wedge \Big(\bigwedge_{w\in W}\bigwedge_{w'\in W}\Boxmulti{N} (\psi_{[w]}\rightarrow \pm \lozenge \psi_{[w']})\Big)\]
where, $\psi_{[w]}\rightarrow \pm \lozenge \psi_{[w']}$ is equal to $\psi_{[w]}\rightarrow \lozenge \psi_{[w']}$ if there are $j\in [w]$ and $j'\in [w']$ such that $(j,j')\in R$, and equal to $\psi_{[w]}\rightarrow \neg \lozenge \psi_{[w']}$ otherwise. 
This $\delta_\A$ has the desired properties. See the proof in the appendix for more details.

Fix now a tabular quasi-normal modal logic $L$. Since $L$ is tabular, there is a finite set of rooted finite frames $\mathcal{F}_L=\{(\F_{1},0),\ldots,(\F_{n},0)\}$ with $L=\text{Log}(\mathcal{F}_{L})$. Fix such a set. Note that we assume without loss of generality that the root of each $\F_i$ is $0$. We will often be sloppy and write
$\F\in\mathcal{F}$  instead of $(\F,0)\in\mathcal{F}$.
Let $N=\max\{|\F|\mid \F\in \mathcal{F}_L\}$. 
By an \emph{abstract $\Phi$-model for $L$}
we will mean an abstract $\Phi$-model whose frame belongs to $\mathcal{F}_L$.
If $\A$ is an abstract $\Phi$-model for $L$, its \emph{$\Phi$-bisimulation class} with respect to $L$, denoted $[\A]_{L}$, is the set of abstract $\Phi$-models $\A'$ for $L$ such that $\A,0\bisim_{\Phi} \A',0$. A formula $\delta\in \text{ML}(\Phi)$ is an \emph{abstract $\Phi$-class identifier for $[\A]_L$} if it defines the class $[\A]_{L}$ in the sense that for every abstract $\Phi$-model $\A'$ for $L$, we have $\A',0\models \delta$ if, and only if, $\A'\in [\A]_{L}$.
We call $\delta$ an \emph{abstract $\Phi$-class identifier for $L$} if there is an abstract $\Phi$-model $\A$ for $L$ such that $\delta$ is an abstract $\Phi$-class identifier for $[\A]_L$. 
If $\delta$ is an abstract $\Phi$-class identifier for $[\A]_L$ we also write $[\delta]_L$ for $[\A]_L$.

Note that the formula $\delta_\A$ from Proposition~\ref{prop:identifier_size} is an abstract $\Phi$-class identifier for $[\A]_{L}$ since $N$ is fixed as $N=\max\{|\F|\mid \F\in \mathcal{F}_L\}$.

\begin{example}\label{ex:identifier}Let $\A_1$ and $\A_2$ be as in Example~\ref{ex:abstract}. The formula $\phi_2\wedge \lozenge \phi_1\wedge\lozenge \phi_3\wedge \square (\phi_1\vee\phi_3)$ is an abstract $\Phi$-class identifier for $[\A_1]_{L_\G}=[\A_2]_{L_\G}$.\end{example}

\begin{definition}
A \emph{$\sigma$-encoding} for $L$ is a pair $(\Gamma,\{\Delta_\Phi\}_{\Phi\in \Gamma})$ where $\Gamma$ is a set of $\sigma$-EMEs and $\Delta_\Phi$ is a set of abstract $\Phi$-class identifiers for $L$ such that for every model $\M$ based on a frame in $\mathcal{F}_L$, there are a $\Phi\in \Gamma$ and $\delta\in \Delta_\Phi$ such that $\Phi$ is a $\sigma$-cover of $\M$ and $\M,0\models \delta$.
\end{definition}

The following proposition 
summarizes the relevant results proved above in a more compact way:

\begin{proposition}
\label{prop:encoding_size}
Fix a tabular quasi-normal modal logic $L$.
Given a signature $\sigma$, we can compute in polynomial
time a $\sigma$-encoding for $L$. 
\end{proposition}
\begin{proof}
We first use Lemma~\ref{lemma:cover_amount} to construct, 
for each $\F\in\mathcal{F}_L$, in time polynomial in $|\sigma|$, a set $\Gamma_\F$ of
$\sigma$-EMEs such that every model based on $\F$ has a $\sigma$-cover in $\Gamma_\F$.  Let $\Gamma=\bigcup_{\F\in\mathcal{F}_L} \Gamma_\F$.
Next, we apply Proposition~\ref{prop:identifier_size} 
to obtain, for each $\Phi\in\Gamma$ and 
for each abstract $\Phi$-model $\A$ for $L$, an
abstract $\Phi$-class identifier $\delta_\A$ for $L$ (recall that we fixed $N=\max\{|\F|\mid \F\in \mathcal{F}_L\}$).
Note that there are at most $\Sigma_{\F\in\mathcal{F}_L}|\Phi|^{|\F|}$ many
abstract $\Phi$-models for $L$, which is polynomial in 
$|\Phi|$ and hence in $|\sigma|$.
Let $\Delta$ be
the set of all these abstract class identifiers.
It follows from the  construction that $(\Gamma,\Delta)$ is a $\sigma$-encoding for $L$.
\end{proof}

\subsection{From Modal Logic to Propositional Logic and Back}
\label{subsec:there_and_back}
In this section, we use propositional formulas to speak about pointed models. Given $\F=(W,R)$ with root $0$ and a set $\sigma$ of atoms, we form the set of atoms $\sigma_{\F}= \sigma \times W$. For notational convenience we denote $(p,w)\in \sigma_{\F}$ as $p_w$. Then $\M=(\F,V)$ defines a propositional valuation $v_{\M}: \sigma\times W \rightarrow \{0,1\}$ for the language PL$(\sigma_{\F})$ by setting 
\[
v_{\M}(p_{w})=
\begin{cases}
1 \text{ if $\M,w\models p$}\\
0 \text{ if $\M,w\models \neg p$}
\end{cases}
\]
Observe that, conversely, every propositional valuation $v$ for PL$(\sigma_{\F})$ defines a modal $\sigma$-model $\M_{v}=(\F,V_{v})$ by setting $\M_{v},w\models p$ iff $v(p_{w})=1$.   
For every modal formula in ML$(\sigma)$ we can easily find an `equivalent' propositional formula in PL$(\sigma_{\F})$ which we define next. 
\begin{definition}
The translation functions $\tr_{\F,w}:\text{ML}(\sigma) \rightarrow \text{PL}(\sigma_{\F})$ with $w\in W$ are defined by mutual induction:
\begin{align*}
\tr_{\F,w}(p)= {}& p_w\\
\tr_{\F,w}(\neg \phi) = {} & \neg \tr_{\F,w}(\phi)\\
\tr_{\F,w}(\phi\wedge \psi) = {} & \tr_{\F,w}(\phi)\wedge \tr_{\F,w}(\psi)\\
\tr_{\F,w}(\lozenge \phi) = {} & \bigvee_{(w,w')\in R}\tr_{\F,w'}(\phi)
\end{align*}
\end{definition}
We summarize the main properties of $\tr_{\F,w}(\phi)$.
\begin{proposition}\label{prop:connect}
Let $\F=(W,R)$ be a finite frame with root $0$, $w\in W$, and $\M$ a model based on $\F$.
\begin{enumerate}
    \item for every $\phi\in \text{ML}(\sigma)$, we have $\M,w\models \phi$ iff $v_{\M}\models \tr_{\F,w}(\phi)$.
    \item for every $\phi\in \text{ML}(\sigma)$, $(\F,0)\models \varphi$ iff $\tr_{\F,0}(\phi)$ is a tautology. 
\end{enumerate}
\end{proposition}
Note that not every propositional formula over $\sigma\times W$ is equivalent to the translation of a modal formula.
\begin{example} Consider the following pointed models $(\M_{3},0)$ and $(\M_{4},0)$:
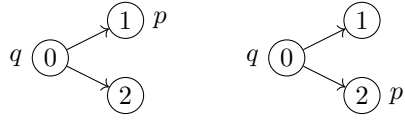
\begin{center}
\begin{tikzpicture}
    \node[draw,circle, inner sep = 2pt,minimum size=8pt, label=left:{$q$}] (s0) at (0,0) {0};
    \node[draw,circle, inner sep = 2pt,minimum size=8pt, label=right:{$p$}] (s1) at (1,0.5) {1};
    \node[draw,circle, inner sep = 2pt,minimum size=8pt, label=right:{}] (s2) at (1,-0.5) {2};

	\draw[->] (s0) -- (s1);
	\draw[->] (s0) -- (s2);
\end{tikzpicture}\hspace{20pt}
\begin{tikzpicture}
    \node[draw,circle, inner sep = 2pt,minimum size=8pt, label=left:{$q$}] (s0) at (0,0) {0};
    \node[draw,circle, inner sep = 2pt,minimum size=8pt, label=right:{}] (s1) at (1,0.5) {1};
    \node[draw,circle, inner sep = 2pt,minimum size=8pt, label=right:{$p$}] (s2) at (1,-0.5) {2};

	\draw[->] (s0) -- (s1);
	\draw[->] (s0) -- (s2);
\end{tikzpicture}
\end{center}
Since these models are $\sigma$-bisimilar for $\sigma=\{p,q\}$, no 
formula in $\text{ML}(\sigma)$ can distinguish between them. It follows that no translation $tr_{\F,0}(\varphi)$
of a formula $\varphi$ in $\text{ML}(\sigma)$ can distinguish between the corresponding propositional models $v_{\M_{3}}$ and $v_{\M_{4}}$.
This implies that no translation $tr_{\F,0}(\varphi)$ of a formula $\varphi$ in $\text{ML}(\sigma)$ is
equivalent to the propositional formula $p_1$, since that formula does distinguish between the two models.
\end{example}

This concludes the presentation of the ``forward'' translation $tr$, i.e., from 
modal formulas to propositional formulas. 
We also need a reverse translation $rt$, 
in order to translate the propositional uniform interpolant back to a modal formula. For this, we make use of the techniques we developed in the previous subsection.
In particular, we make use of Proposition~\ref{prop:encoding_size} and Lemma~\ref{lemma:inv}.

Fix a tabular quasi-normal modal logic $L$, and let $\mathcal{F}_L$ be a finite set of finite rooted frames such that $L=\text{Log}(\mathcal{F}_L)$. Let $N=\max\{|\F'|\mid \F'\in \mathcal{F}_L\}$. 
Recall that we denote the root of each $\F\in\mathcal{F}_L$ by $0$.

\begin{definition}\label{def:inverse}
Fix a tabular quasi-normal modal logic $L$,
and let $\F\in \mathcal{F}_L$.  Then $\rt_{\F,L}:\PL(\sigma_\F)\to \ML(\sigma)$ is given by
\[ \rt_{\F,L}(\xi) = \bigwedge_{\Phi\in \Gamma}\bigwedge_{\delta\in \Delta_\Phi}\left((\inv(\sigma,\Phi)\wedge \delta)\rightarrow \bigvee_{(\F,f)\in [\delta]_L}\xi^f \right)\]
where
 $(\Gamma,\{\Delta_\Phi\}_{\Phi\in \Gamma})$ is a $\sigma$-encoding for $L$ as given by Proposition~\ref{prop:encoding_size}, and 
$\xi^f$ is the result of replacing, in $\xi$, each indexed atom $p_w\in \sigma_{\F}$ by $\Diamondmulti{N}(f(w)\wedge p)$.
\end{definition}

By the results in the previous subsection, all computations involved in this translation can be performed in polynomial time (assuming $L$ is fixed, and hence $N=\max_{\F\in \mathcal{F}_L}|\F|$ can be treated as a constant). In particular, note that the number of $(\F,f)\in[\delta]_L$ is 
at most $|\Phi|^{|\F|}\leq |\Phi|^N$, hence is also polynomial.

\begin{restatable}{proposition}{propbisim}
\label{prop:bisim}
Fix a tabular quasi-normal modal logic $L$. Let $\M,0$ be any pointed model based on a pointed frame in $\mathcal{F}_L$. Furthermore, let $\F\in\mathcal{F}_L$ and let $\xi\in \PL(\sigma_\F)$. Then the following are equivalent:
\begin{enumerate}
\item    
$\M,0\models \rt_{\F,L}(\xi)$,
\item there is a model $\M'$ based on the frame $\F$ such that $\M,0\bisim_{\sigma} \M',0$ and $v_{\M'}\models \xi$.
\end{enumerate}
\end{restatable}

\null 

The proof uses the following lemma, which we prove first.

\begin{lemma}
\label{lemma:bisim_exchange}
Let $\Phi$ be a $\sigma$-cover of $\M$, $Z$ a $\sigma$-bisimulation between $\M$ and $\M'$ such that $w Z w'$ and $\phi\in \Phi$ the formula such that $\M',w'\models \phi$. Then for every $p\in \sigma$, 
$v_{\M'}\models p_{w'}$ if, and only if, $\M,0\models \Diamondmulti{N}(\phi\wedge p)$.
\end{lemma}

\begin{proof}
$(\Rightarrow)$ Suppose $v_{\M'}\models p_{w'}$. Then $\M',w'\models p$. By assumption, $\M',w'\models \phi$, so $\M',w'\models \phi\wedge p$. As $Z$ is a $\sigma$-bisimulation and $\phi\wedge p \in \ML(\sigma)$, this implies that $\M,w\models \phi\wedge p$, and therefore $\M,0\models \Diamondmulti{N}(\phi \wedge p)$. 

$(\Leftarrow)$ Suppose $\M,0\models \Diamondmulti{N}(\phi \wedge p)$. Because $Z$ is a $\sigma$-bisimulation and $\M',w'\models \phi$, we have $\M,w\models \phi$. As $\Phi$ is a $\sigma$-cover of $\M$, the value of $p$ is constant across all worlds that satisfy $\phi$. From $\M,0\models \Diamondmulti{N} (\phi \wedge p)$ it therefore follows that $\M,w\models p$, which implies that $\M',w'\models p$. It follows that $v_{\M'}\models p_{w'}$.
\end{proof}

\begin{proof}[Proof of Proposition~\ref{prop:bisim}]
$(\Rightarrow)$ Suppose $\M,0\models \rt_\F(\xi)$. Because $(\Gamma,\{\Delta_\Phi\}_{\Phi\in \Gamma})$ is a $\sigma$-encoding, there are $\Phi\in \Gamma, \delta\in \Delta_\Phi$ such that $\Phi$ is a $\sigma$-cover of $\M$ and $\M,0\models \delta$. Let $\A$ be the $\Phi$-abstraction of $\M$.
We then have $\M,0\models \inv(\sigma,\Phi)\wedge \delta$, so the antecedent of $\rt_\F(\xi)$ is satisfied for these $\Phi$ and $\delta$. By assumption, $\M,0\models \rt_\F(\xi)$, so the consequent of $\rt_\F(\xi)$ must also hold, i.e., $\M,0\models \bigvee_{(\F,f)\in [\delta]_L}\xi^f$. Let $\A'=(\F,f)\in [\delta]_L$ be such that $\M,0\models \xi^f$.

We are given that $\A'\in [\delta]_L$. Because $\M,0\models\inv(\sigma,\Phi)\wedge\delta$ we also have $\A\in [\delta]_L$. So there is an abstract $\Phi$-bisimulation $Z$ between $\A$ and $\A'$. It follows that there is some $\M'=(\F',V')$ such that $\A'$ is the $\Phi$-abstraction of $\M'$ and $Z$ is a $\sigma$-bisimulation between $\M$ and $\M'$.
Because $\A'=(\F,f)$ is the $\Phi$-abstraction of $\M'$, we have $\M',w'\models f(w')$ for all $w'\in W'$. The conditions of Lemma~\ref{lemma:bisim_exchange} are therefore satisfied, which implies that $\M,0\models \Diamondmulti{N} (f(w')\wedge p)$ if, and only if, $v_{\M'}\models p_{w'}$. From $\M,0\models \xi^f$ it therefore follows that $v_{\M'}\models \xi$. We have seen already that $\M,0$ and $\M',0$ are $\sigma$-bisimilar, so this completes the left-to-right direction of the proof.

$(\Leftarrow)$ Suppose that $\M'=(\F,V')$ is $\sigma$-bisimilar to $\M$ and that $v_{\M'}\models \xi$. 
We must show that
$\M,0\models \rt_\F(\xi)$.
Let $\Phi\in \Gamma$ and $\delta\in \Delta_\Phi$ be such that the antecedent of $\rt_\F(\xi)$ is satisfied, i.e., $\M,0\models \inv(N,\sigma,\Phi)\wedge \delta$.
From $\M,0\models \inv(N,\sigma,\Phi)$ it follows that $\Phi$ is a $\sigma$-cover of $\M$. Let $\A$ be the $\Phi$-abstraction of $\M$. Because $\M,0\models \delta$ we also have $\A,0\models \delta$, so $\A\in [\delta]_L$.

Since $\M,0\bisim_\Phi \M',0$, we have that $\Phi$ is also a $\sigma$-cover of $\M'$,
and 
that $\M',0\models\delta$. Let $\A'=(\F,f)$ be the $\Phi$-abstraction of $\M'$. Then  $\A',0\models \delta$, and therefore $\A'\in [\delta]_L$.
Furthermore, for every $w'\in W'$ we have $\M',w'\models f(w')$. By Lemma~\ref{lemma:bisim_exchange} we now have $v_{\M'}\models p_{w'}$ if, and only if, $\M,0\models \Diamondmulti{N}(f(w')\wedge p)$.
Since, 
by assumption, $v_{\M'}\models \xi$, this implies that $\M,0\models \xi^f$. The consequent of $\rt_\F(\xi)$ is therefore also satisfied in $\M,0$. As this holds for every $\Phi$ and $\delta$ for which the antecedent is satisfied, we have $\M,0\models \rt_\F(\xi)$, which was to be shown.
\end{proof}

\begin{example}
The key observation in the proof of Proposition~\ref{prop:bisim} is that if $\M,0$ and $\M',0$ are $\sigma$-bisimilar, with abstractions $\A=(\F,f)$ and $\A'=(\F,f')$, respectively, then $v_{\M'}\models \xi$ if and only if $\M,0\models \xi^{f'}$. This ensures that if $\xi$ is true in some model $\sigma$-bisimilar to $\M,0$, then at least one of the disjuncts of $\bigvee_{(\F,f)\in [\delta]_L}\xi^f$ is true in $\M,0$.

To illustrate this, let $\M_1, \M_2, \A_1, \A_2$ and $Z$ be as in Example~\ref{ex:abstract}. Furthermore, let $f_2$ be the function such that $\A_2=(\G,f_2)$ and let $\M_2'$ be the model such that $Z$ is a $\sigma$-bisimulation between $\M_1$ and $\M_2'$. Note that $\A_2$ is the $\Phi$-abstraction of both $\M_2$ and $\M_2'$. 

Take $\xi = s_1\in \mathsf{PL}(\sigma_\G)$. Then $\xi$ is not true in $v_{\M_1}$ or $v_{\M_2}$, but true in $v_{\M_2'}$. So $\xi$ is true in some model $\sigma$-bisimilar to $\M_1$, namely $\M_2'$. 

Because $f_2$ is the function associated with the abstraction of $\M_2'$, we then have $\M_1\models \xi^{f_2}$. Indeed, $\xi^{f_2}=\Diamondmulti{N}(s \wedge f_2(w_1))=\Diamondmulti{N}(s \wedge \neg p \wedge \neg r)$, which holds in $\M_1,0$.
\end{example}

The following result implies Theorem~\ref{thm:main} for $L=\text{Log}(\mathcal{F})$ for a singleton $\mathcal{F}=\{(\F,0)\}$.
\begin{theorem}\label{thm:single-frame}
    Fix a tabular quasi-normal modal logic $L$ with
    $L=\text{Log}(\mathcal{F})$ for a singleton $\mathcal{F}=\{(\F,0)\}$. Let $\varphi\in \ML$ and let $\sigma\subseteq \text{sig}(\varphi)$. If 
    $\xi\in \PL(\sigma_{\F})$ is a propositional uniform $\sigma_{\F}$-interpolant for $tr_{\F,0}(\varphi)$, then
    $rt_{\F,L}(\xi)$ is a strongest $L(\sigma)$-implicate of $\varphi$.
\end{theorem}
\begin{proof}
(1) We show that $\varphi \rightarrow rt_{\F,L}(\xi) \in L$. Assume $\M,0\models \varphi$
with $\M$ based on $\F$. Then $v_{\M}\models tr_{\F,0}(\varphi)$. Hence
    $v_{\M}\models \xi$. By the backward direction of Proposition~\ref{prop:bisim} (using the identity bisimulation),
    $\M,0\models rt_{\F,L}(\xi)$. 

(2)  Suppose that
 $\varphi \rightarrow \chi\in L$ for some
 $\chi\in \ML(\sigma)$.
 We must show that
 $\rt_{\F,L}(\xi)\to \chi\in L$.
 Assume $\M,0\models rt_{\F,L}(\xi)$. By Proposition~\ref{prop:bisim}, there is a $\sigma$-bisimilar $\M',0$ based on $\F$ with $v_{\M'}\models \xi$. Next note that by Proposition~\ref{prop:connect}, $tr_{\F,0}(\varphi) \rightarrow tr_{\F,0}(\chi)$ is a tautology. Hence, as $\xi$ is a uniform $\sigma_{\F}$-interpolant for $tr_{\F,0}(\varphi)$, $\xi \rightarrow \tr_{\F,0}(\chi)$ is a tautology. It follows that $v_{\M'} \models tr_{\F,0}(\chi)$ and so $\M',0\models \chi$. Therefore, by invariance under $\sigma$-bisimulations, $\M,0\models\chi$.
\end{proof}
\subsection{From Single Frames to Multiple Frames}
\label{subsec:single_to_multiple}
We now lift Theorem~\ref{thm:single-frame} to tabular quasi-normal logics defined by a \emph{finite set} of finite rooted frames.
Let as before $\mathcal{F}_{L}$ be a finite set of finite pointed frames and $L= \text{Log}(\mathcal{F}_{L})$.
We first show that the strongest $L(\sigma)$-implicate of a modal formula $\varphi$ can be obtained in polynomial time from the propositional uniform $\sigma_{\F}$-interpolants of $tr_{\F}(\varphi)$, $\F\in \mathcal{F}_{L}$. To strengthen this result to a reduction of $L$-implicate computation to a \emph{single} propositional uniform interpolant, we then generalize the translation $tr_{\F}$ to sets $\mathcal{F}_{L}$ of frames and show how uniform interpolants for $tr_{\mathcal{F}}(\varphi)$ are obtained from uniform interpolants for $tr_{\F}(\varphi)$, $\F\in \mathcal{F}_{L}$.
\begin{theorem}\label{thm:many-frame}
    Fix a tabular quasi-normal modal logic $L$ with
    $L=\text{Log}(\mathcal{F}_{L})$. Let $\varphi\in \ML$ and $\sigma\subseteq \text{sig}(\varphi)$. Let $\xi_{\F} \in\PL(\sigma_{\F})$ be propositional uniform $\sigma_{\F}$-interpolants for $tr_{\F}(\varphi)$, for $\F\in \mathcal{F}_{L}$. Then
    $\bigvee_{\F\in \mathcal{F}_{L}}\rt_{\F,L}(\xi_{\F})$ (cf.~Definition~\ref{def:inverse}) is a strongest $L(\sigma)$-implicate of $\varphi$.
\end{theorem}
\begin{proof}
(1) We show that $\varphi \rightarrow \bigvee_{\F\in \mathcal{F}_{L}}\rt_{\F,L}(\xi_{\F})\in L$. Assume $\M,0\models \varphi$
with $\M$ based on $\F\in \mathcal{F}_{L}$. Then $v_{\M}\models tr_{\F}(\varphi)$. Hence
    $v_{\M}\models \xi_{\F}$. By Proposition~\ref{prop:bisim},
    $\M,0\models \rt_{\F,L}(\xi_{\F})$, as required.

(2)  We show that $\bigvee_{\F\in \mathcal{F}_{L}}\rt_{\F,L}(\xi_{\F})\rightarrow \chi \in L$ for all \ML$(\sigma)$-formulas $\chi$ with $\varphi \rightarrow \chi\in L$. To this end, it suffices to show that if $\M,0\models \bigvee_{\F\in \mathcal{F}_{L}}\rt_{\F,L}(\xi_{\F})$ for some $\M$ based on $\F\in \mathcal{F}_{L}$, then there exists a $\sigma$-bisimilar $\M'$ with $\M',0\models \varphi$ based on some $\F'\in \mathcal{F}_{L}$. So assume $\M,0\models \bigvee_{\F\in \mathcal{F}_{L}}\rt_{\F,L}(\xi_{\F})$ with $\M$ based on $\F$. Then we can take $\F'\in \mathcal{F}_{L}$ with $\M,0\models \rt_{\F',L}(\xi_{\F'})$. By Proposition~\ref{prop:bisim}, there is a $\sigma$-bisimilar $\M'$ based on $\F'$ with $v_{\M'}\models \xi_{\F'}$. As $\xi_{\F'}$ is a uniform $\sigma_{\F'}$-interpolant for $tr_{\F'}(\varphi)$, we can modify the values of non-$\sigma_{\F'}$ atoms in $v_{\M'}$ to obtain a model $v$ with $v\models tr_{\F'}(\varphi)$. But then $\M'',0\bisim_{\sigma}\M',0\bisim_{\sigma}\M,0$ and $\M'',0\models \varphi$ for the model $\M''$ based on $\F'$ with $v_{\M''}=v$. 
\end{proof} 
The above theorem can be interpreted as a  Turing reduction: it provides an efficient algorithm for computing a strongest $L(\sigma)$-implicate for a given modal formula, where the algorithm is allowed to ask 
one or more queries to an oracle that computes propositional uniform interpolants. In particular, 
it follows that if propositional 
uniform interpolants can be computed in polynomial time, then strongest $L(\sigma)$-implicates can be computed in polynomial time. Theorem~\ref{thm:main} from the introduction (restated below) is slightly stronger, as it provides an algorithm that asks only one oracle query. Our next aim is therefore to 
strengthen Theorem~\ref{thm:many-frame} to bridge this gap.

\thmmain*

\begin{proof} We only sketch the proof here, see the Appendix for a detailed proof.
First we extend translation $tr_{\F}$ to a translation $tr_{L}$ for any tabular quasi-normal logic $L$. Fix $L=\text{Log}(\mathcal{F}_{L})$ for a finite set $\mathcal{F}_{L}$ of finite rooted frames. We assume that for $(\F,w),(\F',w')\in \mathcal{F}_{L}$ with $(\F,w)\not=(\F',w')$ the sets of worlds in $\F$ and $\F'$ are mutually disjoint (and so do not use $0$ to denote the roots of frames in $\mathcal{F}_{L}$).
Take atoms $r_{\F,w}$, for $(\F,w)\in \mathcal{F}_{L}$. They are used to identify the frame $(\F,w)\in \mathcal{F}_{L}$ in which we evaluate a modal formula. For any
signature $\sigma$, let $\sigma_{L}= \bigcup_{(\F,w)\in \mathcal{F}_{L}}\sigma_{\F,w}$. The signatures  $\sigma_{\F,w}$ are assumed to be mutually disjoint and also disjoint from $\{r_{\F,w}\mid (\F,w)\in \mathcal{F}_{L}\}$. Let $\hat{\sigma}= \sigma_{L}\cup \{r_{\F,w}\mid (\F,w) \in \mathcal{F}_{L}\}$. Next let for any modal formula $\varphi$,
\[
tr_{L}(\varphi)= \text{Unique}(\mathcal{F}_{L})
\wedge \bigwedge_{(\F,w) \in \mathcal{F}_{L}} (r_{\F,w}\rightarrow tr_{\F,w}(\varphi)).
\]
where 
\[\text{Unique}(\mathcal{F}_{L})=(\bigvee_{(\F,w)\in \mathcal{F}_{L}} (r_{\F,w} \wedge \bigwedge_{(\mathfrak{G},v)\in \mathcal{F}_{L}\setminus \{(\F,w)\}} \neg r_{\mathfrak{G},v}))
\]
Hence $tr_{L}(\varphi) \in \PL(\hat{\sigma})$ uses $\text{Unique}(\mathcal{F}_{L})$ to pick $(\F,w)\in \mathcal{F}_{L}$ and then states that $tr_{\F,w}(\varphi)$ holds. 

\medskip

We next define the translation $rt_{L}$ (which is meaningful only if applied to  propositional uniform interpolants). To this end, we first construct from a propositional uniform $\hat{\sigma}$-interpolant of $tr_{L}(\varphi)$ propositional uniform $\sigma_{\F,w}$-interpolants for $tr_{\F,w}(\varphi)$, for every $(\F,w)\in \mathcal{F}_{L}$. Assume $\varphi$ is given and $\sigma\subseteq \text{sig}(\varphi)$. Let $\xi$ be any propositional uniform $\hat{\sigma}$-interpolant of $tr_{L}(\varphi)$. Then obtain for $(\F,w)\in \mathcal{F}_{L}$ the formula $\xi^{\uparrow\F,w}$ from $\xi$ by
\begin{itemize}
    \item replacing $r_{\F,w}$ by $\top$,
    \item replacing all $r_{\mathfrak{G},v}$ with $(\mathfrak{G},v)\in \mathcal{F}_{L}\setminus\{(\F,w)\}$ by $\bot$, and
    \item replacing all $p_{\mathfrak{G},v}$ with $v$ not in $(\F,w)$ by $\bot$.
\end{itemize}
It is easy to see that $\xi^{\uparrow\F,w}$ is a propositional uniform $\sigma_{\F,w}$-interpolant for $tr_{\F,w}(\varphi)$, for every $(\F,w)\in \mathcal{F}_{L}$. Now we simply define for any $\xi\in \PL(\hat{\sigma})$, 
\[
rt_{L}(\xi) = \bigvee_{(\F,w) \in \mathcal{F}_{L}}rt_{(\F,w),L}(\xi^{\uparrow\F,w})
\]
and show using Theorem~\ref{thm:many-frame} that $rt_{L}$ is as required.
\end{proof}
\section{Constructing Craig Interpolants via Propositional Logic}
\label{sec:CraigInt}
We show that if a tabular quasi-normal logic $L=\text{Log}(\mathcal{F}_{L})$ has CIP, then one can compute Craig interpolants for an implication $\varphi\rightarrow \psi\in L$ from propositional Craig interpolants for $tr_{L}(\varphi) \rightarrow tr_{L}(\psi)$ in polynomial time. To this end, we first give a criterion for CIP for tabular quasi-normal logics which shows that $L$ has CIP iff no non-trivial bisimulations between models based on frames in $\mathcal{F}_{L}$ exist. We then show
that the backward translation $\bigvee_{\F\in \mathcal{F}_{L}}rt_{\F,L}(\xi_{\F})$ is a Craig interpolant for $\varphi\rightarrow \psi$ if the $\xi_{\F}$ are Craig interpolants
for $tr_{\F}(\varphi)\rightarrow tr_{\F}(\psi)$ in Log$(\F)$ for $\F\in \mathcal{F}_{L}$. 

We close with a discussion of Craig interpolants for modal logics that do not have CIP.

Let $L=\text{Log}(\mathcal{F}_{L})$ with $\mathcal{F}_{L} =\{(\F_{1},0),\ldots,(\F_{n},0)\}$. Clearly, we may assume that $\mathcal{F}_{L}$ is \emph{reduced}
in the sense that $\text{Log}(\F_{i},0)\not\subseteq \text{Log}(\F_{j},0)$ for $i\not=j$. We start by giving a criterion for CIP of $L$. The \emph{$\sigma$-reduct} of a model $\M$ is the restriction of $\M$ to atoms in $\sigma$. An \emph{isomorphism} between the $\sigma$-reducts of pointed models $\M_{1},w_{1}$ and $\M_{2},w_{2}$ is an isomorphism $f$ between the underlying frames with $f(w_{1})=w_{2}$ such that $\M_{1},w\models p$ iff $\M_{2},f(w)\models p$, for all $p\in \sigma$.
\begin{restatable}{theorem}{thmbsimCIP}
\label{thm:bisimCIP}
Let $L=\text{Log}(\mathcal{F}_{L})$ be a tabular quasi-normal logic with $\mathcal{F}_{L}$ a reduced set of finite rooted frames. Then $L$ has CIP if, and only if, for all signatures $\sigma$ the following holds:
    if $\M_{1},0 \bisim_{\sigma} \M_{2},0$ with $\M_{1}$ a model based on $\F_{i}\in \mathcal{F}_{L}$ and $\M_{2}$ a model based on $\F_{j}\in \mathcal{F}_{L}$, then the $\sigma$-reducts of $\M_{1},0$ and $\M_{2},0$ are isomorphic and, in particular, $i=j$.   
\end{restatable}
\begin{proof}
    We use the following criterion for CIP for tabular modal logics which follows immediately from known sufficient conditions~\cite{DBLP:conf/amast/Marx98}:

    \medskip
    \noindent
    \emph{Fact 1}. $L$ has CIP iff for all modal formulas $\varphi,\psi$ and $\sigma=\text{sig}(\varphi)\cap\text{sig}(\psi)$ the following holds: if there are models $\M_{1},0\models \varphi$ and $\M_{2},0\models \psi$ with $\M_{1},0 \bisim_{\sigma} \M_{2},0$ such that $\M_{1}$ is based on $\F_{i}\in \mathcal{F}_{L}$ and $\M_{2}$ is based on $\F_{j}\in \mathcal{F}_{L}$, then there is a model $\M',0$ based on some $\F_{k}\in \mathcal{F}_{L}$ with $\M',0\models \varphi\wedge \psi$.

    \medskip
    
    The direction from right to left is a direct consequence of Fact~1. 
    We now show the converse direction. Assume $(\F,0)$ is a finite rooted frame and $N=\max_{\F\in \mathcal{F}_L}|\F|$. Define the \emph{diagram} diag$(\mathfrak{F},0)$ of $(\F,0)$ (also known as the splitting-, Jankov-, or canonical formula of $(\F,0)$~\cite{DBLP:books/el/07/WolterZ07}) as the conjunction of the following formulas (with $q_{w}$ fresh atoms for $w$ in $\F$):
    \begin{description}
        \item $q_{0} \wedge \Box^{\leq N}\bigvee_{v\in W} q_{v}$
        \item $\Box^{\leq N}(q_{v}\rightarrow \neg q_{v'})$ if $v\not= v'$
        \item $\Box^{\leq N} (q_{v} \rightarrow \Diamond q_{v'})$ if $vRv'$
        \item $\Box^{\leq N} (q_{v} \rightarrow \neg\Diamond q_{v'})$ if not $vRv'$.
    \end{description}
    The following facts about diagrams are straightforward:

    \medskip
    \noindent
    \emph{Fact 2.} (1) If there exists a model $\M$ based on $\mathfrak{F}_{j}\in \mathcal{F}_{L}$ with $\M,0\models \text{diag}(\mathfrak{F}_{i},0)$, then Log$(\F_{i},0)\supseteq \text{Log}(\F_{j},0)$ and so $i=j$.

    (2) If $\M,0\models \text{diag}(\mathfrak{F}_{i},0)$ for a model $\M$ based on $\F_{i}\in \mathcal{F}_{L}$, then for every $w$ in $\F_{i}$ there is exactly one $f(w)$ in $\F_{i}$ with $\M,f(w)\models q_{w}$. $f$ is an automorphism on $(\F_{i},0)$.

    \medskip
    
    Now assume that $L$ has CIP. Assume a signature $\sigma$ and
    $\M_{1},0 \bisim_{\sigma} \M_{2},0$ with $\M_{1}$ a model based on $\F_{i}\in \mathcal{F}_{L}$ and $\M_{2}$ a model based on $\F_{j}\in \mathcal{F}_{L}$ are given. We take disjoint sets of atoms $q_{w,1}$ for $w$ in $\F_{i}$ and $q_{w,2}$ for $w$ in $\F_{j}$. Denote by diag$_{1}$ the diagram
    of $(\F_{i},0)$ using the atoms $q_{w,1}$ and by diag$_{2}$ the diagram of $(\F_{j},0)$ using the atoms $q_{w,2}$, respectively. Note that $i=j$ is allowed. We also assume that the atoms of these diagrams are disjoint from $\sigma$. We now define descriptions $\delta_{\sigma}(\M_{k},0)$ of the $\sigma$-reducts of $\M_{k}$ by setting
    for $L_{\sigma}=\{p,\neg p\mid p\in \sigma\}$ and $k=1,2$: 
    \[
    \delta_{\sigma}(\M_{k},0)=\text{diag}_{k} \wedge \Box^{\leq N}(\bigwedge_{\substack{\M_{k},w\models l \\ l\in L_{\sigma}}}(q_{w,k}\rightarrow l)).
    \]
    It should be clear that $\M_{k},0\models \delta_{\sigma}(\M_{k},0)$ for $k=1,2$. Since $L$ has CIP we can apply Fact 1 because  
    $\sigma=\text{sig}(\delta_{\sigma}(\M_{1},0))\cap \text{sig}(\delta_{\sigma}(\M_{2},0))$. Hence, there is a model $\M'$ based on a $\F_{k}\in \mathcal{F}_{L}$ such that $\M'\models \delta_{\sigma}(\M_{1},0)\wedge \delta_{\sigma}(\M_{2},0)$. By Fact 2 (1), $i=j=k$. 

    By Fact~2 (2), we have automorphisms $f_{1}:(\F_{i},0) \rightarrow (\F_{i},0)$ and $f_{2}:(\F_{i},0) \rightarrow (\F_{i},0)$ with $\M',f_{1}(w)\models q_{w,1}$ and $\M',f_{2}(w)\models q_{w,2}$ for all $w$ in $\F_{i}$. $f_{1}$ and $f_{2}$ are invariant for $\sigma$-atoms in the sense that for all $w$ in $\F_{i}$ and $p\in \sigma$: $\M_{k},w\models p$ iff $\M',f_{k}(w)\models p$, for $k=1,2$. It follows that $f_{1}\circ f_{2}^{-1}$ is an isomorphism between the $\sigma$-reducts of $(\M_{1},0)$ and $(\M_{2},0)$, as required.
\end{proof} 
The following example illustrates Theorem~\ref{thm:bisimCIP}.
\begin{example}\label{example:cip5}
(1) Let $\G$ and $L_{\G}$ be from Example~\ref{example:first}. Observe that for $\sigma=\{p\}$
and $\M_{1}$ and $\M_{2}$ based on $\G$ with $p$ true in $1$ in $\M_{1}$ and $p$ true in $1,2$ in $\M_{2}$ we have $\M_{1},0\bisim_{\sigma}\M_{2},0$ but $\M_{1},0$ and $\M_{2},0$ are not isomorphic. Hence $L_{\G}$ does not have CIP. 

(2) Observe that any two models $\M_{1},0$ and $\M_{2},0$ based on reflexive frames are $\sigma$-bisimilar, for $\sigma=\emptyset$. Hence, if $L=Log(\mathcal{F})$ for a finite set of finite reflexive pointed frames and $L$ has CIP, then $L=Log(\F)$ for a single finite pointed frame $\F$. 
\end{example}
We are now in a position to prove the following variant of Theorem~\ref{thm:many-frame} for Craig interpolant computation.
\begin{theorem}\label{thm:many-frame-Craig}
    Fix a tabular quasi-normal modal logic $L$ with CIP and with
    $L=\text{Log}(\mathcal{F}_{L})$. Let $\varphi\rightarrow \psi\in L$. If 
    $\xi_{\F}\in \PL(\sigma_{\F})$ are propositional Craig interpolants for $tr_\F(\varphi)\rightarrow tr_{\F}(\psi)$, $\F\in \mathcal{F}_{L}$, then
    $\bigvee_{\F\in \mathcal{F}_{L}}rt_{\F,L}(\xi_{\F})$ is a Craig interpolant for $\varphi\rightarrow\psi$ in $L$.
\end{theorem}
\begin{proof}
    Assume $\varphi\rightarrow \psi\in L$ is given. We show that 
    (i) $\varphi \rightarrow \bigvee_{\F\in \mathcal{F}_{L}}rt_{\F,L}(\xi_{\F})\in L$ and (ii) $\bigvee_{\F\in \mathcal{F}_{L}}rt_{\F,L}(\xi_{\F})\rightarrow \psi \in L$. 
    (i) can be shown in exactly the same 
    way as in the proof of Theorem~\ref{thm:many-frame}. To prove (ii), assume
    $\M,0\models \bigvee_{\F\in \mathcal{F}_{L}}rt_{\F,L}(\xi_{\F})$ with $\M$ based on $\F_{i}$. We show that
    $\M,0\models \psi$. Take $\F_{j}$ with $\M,0\models rt_{\F_{j},L}(\xi_{\F_{j}})$.
    By Proposition~\ref{prop:bisim}, there is a $\sigma$-bisimilar $\M',0$ based on $\F_{j}$ with $v_{\M'}\models \xi_{\F_{j}}$. By Theorem~\ref{thm:bisimCIP}, $i=j$ and we have an isomorphism $f$ between the $\sigma$-reducts of $(\M,0)$ and $(\M',0)$.
    We modify $\M'$ in the obvious way to obtain a model $\M''$ such that $f$ is an isomorphism from $(\M,0)$ onto $(\M'',0)$. Then $v_{\M''}\models \xi_{\F_{i}}$ since
    $v_{\M'}$ and $v_{\M''}$ coincide on $\sigma_{\F_{i}}$. We obtain $v_{\M''}\models tr_{\F_{i}}(\psi)$ since $\xi_{\F_{i}}$ is a Craig interpolant for $tr_{\F_{i}}(\varphi) \rightarrow tr_{\F_{i}}(\psi)$. Hence $\M'',0\models \psi$. But then $\M,0\models \psi$ since $(\M,0)$ and $(\M'',0)$ are isomorphic.
\end{proof}
Similarly to the computation of strongest implicates, also Theorem~\ref{thm:many-frame-Craig} can be strengthened 
by showing that to compute modal Craig interpolants only one query to an oracle computing propositional 
Craig interpolants is needed.
\thmmainCIP*
The proof is similar to the proof of Theorem~\ref{thm:main} (now using Theorem~\ref{thm:many-frame-Craig}) and is therefore omitted.

\medskip

We briefly discuss what happens for logics $L$ without CIP. Clearly, in this case there must be $\varphi\rightarrow \psi\in L$ such that for any propositional Craig interpolant $\xi$ for $tr_{L}(\varphi)\rightarrow tr_{L}(\psi)$ the modal formula $rt_{L}(\xi)$ is not a Craig interpolant for $\varphi,\psi$ in $L$. For instance, consider $L_{\G}=\text{Log}(\{(\G,0)\})$ defined in Example~\ref{example:first}. Take the formulas
\[
\gamma_{0} =(\lozenge (p\wedge q)\wedge \lozenge (p\wedge \neg q)), \quad \gamma_{1}= (\lozenge (\neg p\wedge r)\rightarrow \square (\neg p\rightarrow r))
\]
and let $\sigma=\{p\}$. Then the propositional formula $\xi$ stating that ``at least two successors satisfy $p$'' is (up to logical equivalence) the propositional uniform $\sigma_{\G}$-interpolant for $tr_{\G,0}(\gamma_{0})$ and also the unique (up to logical equivalence) Craig interpolant for $tr_{\G,0}(\gamma_{0}) \rightarrow tr_{\G,0}(\gamma_{1})$. The backward translation $rt_{\G,0}(\xi)$ is equivalent to $\Diamond p$. It is a strongest $L_{\G}(\sigma)$-implicate of $\gamma_{0}$ but neither a uniform $L_{\G}(\sigma)$-interpolant for $\gamma_{0}$ nor a Craig interpolant for $\gamma_{0}\rightarrow \gamma_{1}$ in $L_{\G}$.

There has recently been work on the computation of Craig interpolants for logics without CIP~\cite{DBLP:journals/tocl/ArtaleJMOW23,DBLP:journals/corr/abs-2312-05929}. Even for a tabular logic without CIP, it is of interest to compute, given $\varphi\rightarrow\psi\in L$ such that there exists a Craig interpolant for $\varphi,\psi$ in $L$, such a Craig interpolant from any Craig interpolant for $tr_{L}(\varphi)\rightarrow tr_{L}(\psi)$ in polynomial time.
We next show that this is \emph{impossible} if there are valid implications in propositional logic for which all Craig interpolants are of superpolynomial size. 
Assume $\varphi_{n}\rightarrow\psi_{n}$, $n\geq 0$, is a family
of propositional formulas such that all Craig interpolants for $\varphi_{n}\rightarrow \psi_{n}$ are of superpolynomial size. 
Now consider again the tabular logic $L_{\G}$.
Let $\sigma_{n}$ be the signature of $\varphi_{n},\psi_{n}$. Let
\[
\varphi_{n}' = \gamma_{0} \wedge \Box \varphi_{n}, \quad \psi_{n}' = \gamma_{1} \vee \Box \psi_{n}
\]
with $\gamma_{0},\gamma_{1}$ defined above. We assume that $\{p,q,r\}$ is disjoint from $\sigma_{n}$. The following statements hold: 
\begin{enumerate}
    \item there are propositional Craig interpolants of polynomial size for $tr_{\G,0}(\varphi_{n}')\rightarrow tr_{\G,0}(\psi_{n}')$: simply take the Craig interpolant for $tr_{\G,0}(\gamma_{0})\rightarrow \tr_{\G,0}(\gamma_{1})$ defined above;
    \item there are Craig interpolants for $\varphi_{n}'\rightarrow \psi_{n}'$ in $L_{\G}$: take $\Box\chi_{n}$ with $\chi_{n}$ a propositional Craig interpolant for $\varphi_{n}\rightarrow \psi_{n}$; and 
    \item Craig interpolants for $\varphi_{n}'\rightarrow \psi_{n}'$ in $L_{\G}$ are of superpolynomial size:
    from Craig interpolants for $\varphi_{n}'\rightarrow \psi_{n}'$ in $L_{\G}$ one can construct in  polynomial time in a straightforward way propositional Craig interpolants for $\varphi_{n}\rightarrow \psi_{n}$.
\end{enumerate}
Of course, this result does not refute the possibility that a different translation can be used to polynomially reduce Craig interpolants for modal logics without CIP to propositional Craig interpolants.
\section{Lower Bounds}
\label{sec:lower}
\begin{figure*}
\caption{The finite rooted frames for the five pre-tabular normal logics containing S4. A boundary around a set of worlds indicates a \emph{cluster} of worlds with $wRw'$ for all $w,w'$ in the cluster. Reflexive and transitive edges are omitted for clarity.}\label{fig:s4ext}
\centering
\begin{subfigure}{0.08\textwidth}
\caption{$\mathcal{F}_{\text{Grz.3}}$}
\centering
\begin{tikzpicture}[xscale=0.8,yscale=-0.8]
    \node[draw,circle, inner sep = 2pt,minimum size=8pt] (s0) at (0,0) {};
    \node[draw,circle, inner sep = 2pt,minimum size=8pt] (s1) at (0,-1) {};
    \node (s3) at (0,-1.8) {$\vdots$};

	\draw[->] (s1) -- (s0);
\end{tikzpicture}
\end{subfigure}
\begin{subfigure}{0.22\textwidth}
\caption{$\mathcal{F}_{\text{S5}}$}
\centering
\begin{tikzpicture}[xscale=0.8,yscale=-0.8]
    \node[draw,circle, inner sep = 2pt,minimum size=8pt] (s0) at (0,0) {};
    \node[draw,circle, inner sep = 2pt,minimum size=8pt] (s1) at (1,0) {};
    \node[draw,circle, inner sep = 2pt,minimum size=8pt] (s2) at (2,0) {};
    \node (s3) at (2.6,0) {$\cdots$};
\node[draw,rounded corners,fit = (s0) (s3)] {};
\end{tikzpicture}
\end{subfigure}
\begin{subfigure}{0.22\textwidth}
\caption{$\mathcal{F}_{1}$}
\centering
\begin{tikzpicture}[xscale=0.8,yscale=-0.8]
    \node[draw,circle, inner sep = 2pt,minimum size=8pt] (s0) at (0,0) {};
    \node[draw,circle, inner sep = 2pt,minimum size=8pt] (s1) at (1,0) {};
    \node[draw,circle, inner sep = 2pt,minimum size=8pt] (s2) at (2,0) {};
    \node (s3) at (2.6,0) {$\cdots$};
    \node[draw,rounded corners,fit = (s0) (s3)] (c1) {};
    \node[draw,circle, inner sep = 2pt,minimum size=8pt,below of=c1] (s4) {};
    
    \draw[->] (c1) -- (s4);
\end{tikzpicture}
\end{subfigure}
\begin{subfigure}{0.22\textwidth}
\caption{$\mathcal{F}_{2}$}
\centering
\begin{tikzpicture}[xscale=0.8,yscale=-0.8]
    \node[draw,circle, inner sep = 2pt,minimum size=8pt] (s0) at (0,0) {};
    \node[draw,circle, inner sep = 2pt,minimum size=8pt] (s1) at (-1.2,1) {};
    \node[draw,circle, inner sep = 2pt,minimum size=8pt] (s2) at (-0.2,1) {};
    \node[draw,circle, inner sep = 2pt,minimum size=8pt] (s3) at (0.8,1) {};
    \node (s4) at (1.4,1) {$\cdots$};
    
    \draw[->] (s0) -- (s1);
    \draw[->] (s0) -- (s2);
    \draw[->] (s0) -- (s3);
\end{tikzpicture}
\end{subfigure}
\begin{subfigure}{0.20\textwidth}
\caption{$\mathcal{F}_{3}$}
\centering
\begin{tikzpicture}[xscale=0.8,yscale=-0.8]
    \node[draw,circle, inner sep = 2pt,minimum size=8pt] (s0) at (0,0) {};
    \node[draw,circle, inner sep = 2pt,minimum size=8pt] (s1) at (-1.2,1) {};
    \node[draw,circle, inner sep = 2pt,minimum size=8pt] (s2) at (-0.2,1) {};
    \node[draw,circle, inner sep = 2pt,minimum size=8pt] (s3) at (0.8,1) {};
    \node (s4) at (1.4,1) {$\cdots$};
    \node[draw,circle, inner sep = 2pt,minimum size=8pt] (s5) at (0,2) {};
    
    \draw[->] (s0) -- (s1);
    \draw[->] (s0) -- (s2);
    \draw[->] (s0) -- (s3);
    \draw[->] (s1) -- (s5);
    \draw[->] (s2) -- (s5);
    \draw[->] (s3) -- (s5);
\end{tikzpicture}
\end{subfigure}
\end{figure*}
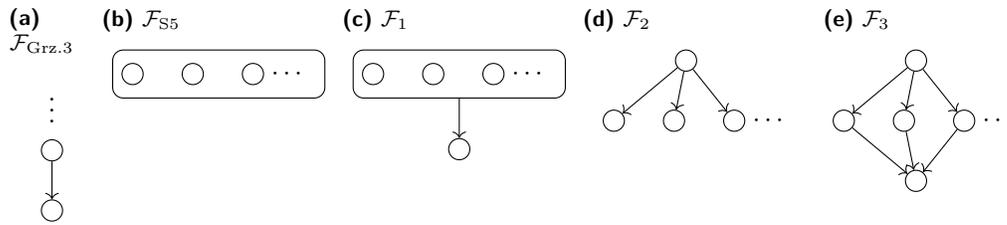

Our aim in this section is to prove exponential lower bounds on the size of strongest implicates and Craig interpolants, for large classes of non-tabular modal logics. Our main result is as follows.

\thmmainlower*

While for Craig interpolants this theorem fully captures our results, for strongest implicates Part~1 follows from a significantly stronger result
based on sufficient model-theoretic conditions for exponential size implicates. These conditions are defined next. 
Recall that a quasi-normal modal logic $L$ has the \emph{poly-size model property} if for every formula $\varphi\not\in L$ there exists a rooted frame $(\F,w)$ validating $L$ and refuting $\varphi$ which is of polynomial size in $|\varphi|$. We say that $L$ has the \emph{exponential growth property} if there is a polynomial function $f$ such that for every $n>0$ there exists a rooted frame $(\F,w)$ validating $L$ such that at least $2^{n}$ worlds are reachable from $w$ by an $R$-path of length at most $f(n)$ starting at $w$. A few standard modal logics such as S5, GL.3, and Grz.3 have both the poly-size model property and exponential growth property. We are in a position now to formulate a sufficient condition for exponential size strongest implicates.
\begin{theorem}\label{thm:lowerimplicate1}
Let $L$ be a 
quasi-normal modal logic contained in a quasi-normal modal logic $L'$ with the poly-size model property and the exponential growth property. Then there exist modal formulas $\varphi_{n}$ and signatures $\sigma_{n}\subseteq \text{sig}(\varphi_{n})$ with $|\varphi_{n}|$ polynomial in $n$
such that the strongest $L(\sigma_{n})$-implicates of $\varphi_{n}$ exist and are of size 
$\geq 2^{n}$.
\end{theorem}
\begin{proof}
    By the poly-size model property of $L'$, we can take $k>0$ such that satisfiable formulas of size $m$ are satisfied in models of size at most $m^{k}$, for sufficiently large $m$. We also take a polynomial function $f$ witnessing the exponential growth property of $L'$. Consider atoms $\sigma_{n}=\{p_{0},\ldots,p_{kn-1}\}$ and let $q_{0},\ldots,q_{kn-1}$ be a set of atoms disjoint from $\sigma_{n}$. Let 
    \begin{align*}
    \varphi_{n}  = & \;\; \Box^{\leq f(kn)}\bigvee_{i=0}^{kn-1}\neg(p_{i}\leftrightarrow q_{i}) \wedge \\ & \;\;\Box^{\leq f(kn)-1} \bigwedge_{i=0}^{kn-1} ((q_i\to \Box q_i)\land (\neg q_i\to \Box \neg q_i))
    \end{align*}
    The formula $\chi_{n}=\bigvee_{t\in T_{kn}}\Box^{\leq f(kn)}\neg t$ is then easily seen to be a strongest $L(\sigma_n)$-implicate of $\varphi_{n}$, 
    where $T_{kn}$ 
    is the set of types $\bigwedge_{i=0}^{kn-1} l_{i}$ with $l_{i} \in \{p_{i},\neg p_{i}\}$ for $i=0,\ldots,kn-1$.
    Assume now for a proof by contradiction that $\chi_{n}$ is equivalent on frames validating $L$ 
    to a formula $\psi_{n}$ of size $<2^{n}$, for sufficiently large $n$. Then this is also the case for $L'\supseteq L$. Take $\neg \psi_{n}$ which states that all $2^{kn}$ types in $T_{kn}$ are
    satisfied in worlds reachable in $f(kn)$ steps. By the exponential growth property of
    $L'$, $\neg\psi_{n}$ is satisfiable 
    in an $L'$-frame. We derive a contradiction since then
    $\neg \psi_{n}$ is satisfied in a model of size $<2^{kn}$, for sufficiently large $n$.    
\end{proof}
We next spell out the consequences of Theorem~\ref{thm:lowerimplicate1} and show that the first part of Theorem~\ref{thm:main-lower} follows.
Observe first that every non-tabular quasi-normal modal logic containing K4 has the exponential growth property since every world reachable from a world is reachable in one step already. Hence, above K4, it suffices to analyze the poly-size model property. Call a normal modal logic $L$ \emph{pre-tabular} if every normal modal logic properly containing $L$ is tabular.\footnote{We use the term pre-tabular logic only in the class of \emph{normal} modal logics. Pre-tabular logics have also been studied in the class of quasi-normal modal logics~\cite{DBLP:books/daglib/0030819}. The two notions are indeed different. We focus on normal ones for simplicity here.} 
Pre-tabular logics have been investigated extensively in modal logic as they separate tabular from non-tabular logics. For the following result and further details on pre-tabular logics, we refer the reader to~\cite[Chapter~12]{DBLP:books/daglib/0030819}.

\begin{figure}
\centering
\caption{The finite rooted frames for the pre-tabular normal logics containing GL. Transitive edges are omitted for clarity.}
\label{fig:GLext}
\begin{subfigure}{0.1\textwidth}
\caption{$\mathcal{F}_{\text{GL.3}}$}
\centering
\begin{tikzpicture}
    \node[draw,circle, inner sep = 2pt,minimum size=8pt] (s0) at (0,0) {};
    \node[draw,circle, inner sep = 2pt,minimum size=8pt] (s1) at (0,-1) {};
    \node[draw,circle, inner sep = 2pt,minimum size=8pt] (s2) at (0,-2) {};
    \node (s3) at (0,-2.4) {$\vdots$};

	\draw[->] (s2) -- (s1);
	\draw[->] (s1) -- (s0);
\end{tikzpicture}
\end{subfigure}\hspace{30pt}
\begin{subfigure}{0.3\textwidth}
\caption{$\mathcal{F}_{n,m}$}
\centering
\begin{tikzpicture}
    \node[draw,circle, inner sep = 2pt,minimum size=8pt] (s1) at (0,0) {};
    \node[draw,circle, inner sep = 2pt,minimum size=8pt] (s2) at (0,-1) {};
    \node (s3) at (0,-1.4) {$\vdots$};
    \node[draw,circle, inner sep = 2pt,minimum size=8pt] (sm) at (0,-2) {};  
    \node[draw,circle, inner sep = 2pt,minimum size=8pt] (tn) at (0,-3) {};  
    \node (t2) at (0,-3.4) {$\vdots$};
    \node[draw,circle, inner sep = 2pt,minimum size=8pt] (t1) at (0,-4) {};
    \node[draw,circle, inner sep = 2pt,minimum size=8pt] (t0) at (0,-5) {};

    \node[draw,circle, inner sep = 2pt,minimum size=8pt] (u0) at (1,-4) {};
    \node[draw,circle, inner sep = 2pt,minimum size=8pt] (u1) at (2,-4) {};
    \node (u3) at (2.6,-4) {$\cdots$};

	\draw[->] (s2) -- (s1);
    \draw[->] (tn) -- (sm);
	\draw[->] (t0) -- (t1);
    \draw[->] (t0) -- (u0);
    \draw[->] (t0) -- (u1);
    \draw[->] (u0) -- (sm);
    \draw[->] (u1) -- (sm);

    \draw [decorate,decoration={brace,amplitude=5pt,raise=10pt}] (0,-4.2) -- (0,-2.8) node[midway,left=15pt]{$n$};
    \draw [decorate,decoration={brace,amplitude=5pt,raise=10pt}] (0,-2.2) -- (0,.2) node[midway,left=15pt]{$m$};
\end{tikzpicture}
\end{subfigure}
\end{figure}
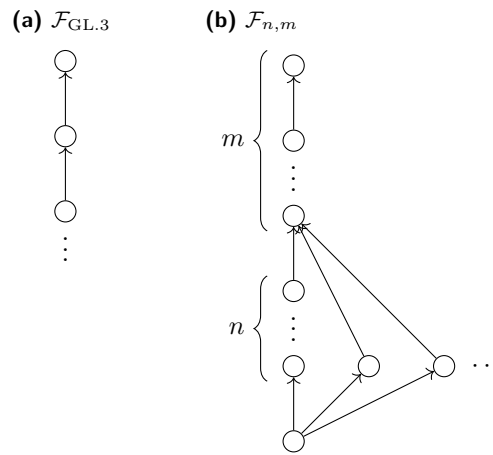

\begin{proposition}  
\label{prop:pretabular}
(1) Every non-tabular normal modal logic is contained in a pre-tabular normal modal  logic.

(2) All normal modal logics containing a tabular normal modal logic are tabular.

(3) There are five pre-tabular normal modal logics containing S4. These are $L(\mathcal{F})$
for $\mathcal{F}$ depicted in Figure~\ref{fig:s4ext}. 

(4) There are countably many pre-tabular normal modal logics containing GL. These are $L(\mathcal{F})$ for $\mathcal{F}$ depicted in Figure~\ref{fig:GLext}.
\end{proposition} 

Observe that $\text{Grz}.3=L(\mathcal{F}_{\text{Grz}.3})$, $\text{S5}=L(\mathcal{F}_{\text{S5}})$ are pre-tabular logics containing S4 and $\text{GL}.3=L(\mathcal{F}_{\text{GL}.3})$ is a pre-tabular logic containing GL. Now the first half of Theorem~\ref{thm:main-lower} follows from Theorem~\ref{thm:lowerimplicate1}, Proposition~\ref{prop:pretabular} (1), and the following result.
\begin{proposition}\label{prop:pretabular-good}
    All pre-tabular normal modal logics containing S4 or GL have both the poly-size model property and, trivially, the exponential growth property.
\end{proposition}
The proof of Proposition~\ref{prop:pretabular-good} is straightforward, given the description of frames for pre-tabular normal modal logics in Figures~\ref{fig:s4ext} and \ref{fig:GLext}. In fact, to construct
poly-size satisfying models one can use selective filtration in the same way as in the
proof of the second half of Theorem~\ref{thm:main-lower} below.

\medskip

We come to the proof of Theorem~\ref{thm:main-lower} (2). We define modal formulas $(\phi_n)_{n=1,2,\ldots}$ and 
 $(\psi_n)_{n=1,2,\ldots}$
 of size polynomial in $n$ and show for  
each pre-tabular normal modal logic $L'$ containing S4 or GL separately that for every normal modal logic $L$ contained in $L'$ Craig interpolants for $\phi_{n}\to\psi_{n}$ exist but are of exponential size.

The proof requires slightly different arguments for each pre-tabular normal modal logic. Consider $\sigma_{n}=\{r\} \cup \{p_{0},\ldots,p_{n-1}\}$ and let $T_{n}$ denote the set of types $\bigwedge_{i=0}^{n-1} l_{i}$ with $l_{i} \in \{p_{i},\neg p_{i}\}$ for $i=0,\ldots,n-1$.
Let $q_{0},\ldots,q_{n-1},q_{0}',\ldots,q_{n-1}'$ be a set of atoms disjoint from $\sigma_{n}$. We aim to enforce the Craig interpolants $
I_{n}=\bigvee_{t\in T_{n}}(r \wedge t \wedge \Diamond (\neg r \wedge t))$.
To this end, define
\[
\chi(p_{0},\ldots,p_{n-1},q_{0},\ldots,q_{n-1}) = \bigwedge_{0\leq i<n}\big((p_{i} \leftrightarrow q_{i}) \wedge (q_{i} \rightarrow \Box q_{i}) \wedge (\neg q_{i}\rightarrow  \Box\neg q_{i})\big)
\]
and obtain $\chi(p_{0},\ldots,p_{n-1},q_{0}',\ldots,q_{n-1}')$ from $\chi(p_{0},\ldots,p_{n-1},q_{0},\ldots,q_{n-1})$ by replacing $q_{i}$ by $q_{i}'$, for all $i<n$.
Then let
\[
\varphi_{n} = r \wedge \chi(p_{0},\ldots,p_{n-1},q_{0},\ldots,q_{n-1}) \wedge \Diamond(\neg r \wedge \bigwedge_{0\leq i<n}(p_{i} \leftrightarrow q_{i}))
\]
and
\[
\psi_{n} = r \wedge (\chi(p_{0},\ldots,p_{n-1},q_{0}',\ldots,q_{n-1}') \rightarrow \Diamond(\neg r \wedge \bigwedge_{0\leq i<n}(p_{i} \leftrightarrow q_{i}')))
\]
Observe that for every pointed frame $\F,0$ we have $\F,0 \models \varphi_{n} \rightarrow \psi_{n}$ and $I_{n}$ is a Craig interpolant for $\varphi_{n}\rightarrow\psi_{n}$ for every quasi-normal modal logic. The following observation follows directly from the fact that every model of $I_{n}$ can be expanded to a model of $\varphi_{n}$ and every model of $\neg I_{n}$ can be expanded to a model of $\neg \psi_{n}$.

\medskip
\noindent
\emph{Fact 1.} If $L$ is a quasi-normal modal logic, then every Craig interpolant for $\varphi_{n}\rightarrow \psi_{n}$ is equivalent to $I_{n}$ in $L$.

\medskip

We now argue that for every $L$ contained in a normal pre-tabular logic containing S4
or GL, there is no formula equivalent to $I_{n}$ on frames validating $L$ of size $<2^{n}$.
We show this for $L \subseteq Log(\mathcal{F}_{\text{Grz.3}})$ and for $L \subseteq Log(\mathcal{F}_{\text{S5}})$, the remaining pre-tabular logics are considered in the appendix. 

Given any transitive model $\M=(\F,R,V)$, a world $w\in W$
is called \emph{maximal for a formula $\chi$ in $\M$} if $\M,w\models \chi$ and for any $w'$ with $\M,w'\models\chi$ and $wRw'$ it follows that $w'=w$ or $w'Rw$. 

\medskip

\noindent
Case 1. Let $L \subseteq Log(\mathcal{F}_{\text{Grz.3}})$. 
Assume for a proof by contradiction that $\chi_{n}$ is equivalent to $I_{n}$ in $L$ and
of size $<2^{n}$.

We construct a model $\M_{1,n}$ based on $\F_{1,n}=(W_{1,n},R_{1,n})$ as follows:
let $W_{1,n}= M_{1,n}\cup N_{1,n}$ with $M_{1,n}= \{a_{0},\ldots,a_{2^{n}-1}\}$ and $N_{1,n}=\{b_{0},\ldots,b_{2^n-1}\}$. Define $R_{1,n}$ as the reflexive and transitive closure of 
\[
\{(a_{i},b_{0})\mid i < 2^{n}\} \cup \{(b_{i},b_{i+1})\} \mid i< 2^{n}-1\}.
\]
Then $\F_{1,n}$ validates $L$. Let $\M_{1,n}$ be defined in such a way that
\begin{itemize}
    \item every type $t\in T_{n}$ is satisfied in exactly one $a_{i}$, $i<2^{n}$;
    \item $r$ is satisfied in $a_{i}$ for all $i<2^{n}$ and not in any $b_{i}$ with $i<2^{n}$;
    \item every type $t\in T_{n}$ is satisfied in exactly one $b_{i}$, $i<2^{n}$.
\end{itemize}
We have $\M_{1,n},a_{i}\models I_{n}$ for all $i<2^{n}$. Hence we have $\M_{1,n},a_{i}\models\chi_{n}$.
Pick for every $\xi\in \text{sub}(\chi_{n})$ satisfied in some world in $N_{1,n}$ a maximal world $b_{\xi}$ in $N_{1,n}$ satisfying $\xi$ in $\M_{1,n}$. For each $i<2^{n}$, let $\M_{i}$ denote the restriction of $\M_{1,n}$ to $\{a_{i}\}\cup \{ b_{\xi}\mid b_{\xi} \in \text{sub}(\chi_{n}) \text{ picked}\}$. Then each pointed $\M_{i},c$ with $c$ in $\M_{i}$ satisfies the same type $t\in T_{n}$ as $\M_{1,n},c$. Moreover, $\M_{i},a_{i}\models \chi_{n}$ for all $i<2^{n}$ can be shown by induction. Hence $\M_{i},a_{i}\models I_{n}$ since the $\M_{i},a_{i}$ are based on frames validating $L$. However, note that there is a type in $T_{n}$ that is not satisfied in any $b_{\xi}$ since the number of $b_{\xi}$ in $\M_{i}$ is $<2^{n}$. Hence there is an $a_{i}$ with $\M_{i},a_{i}\not\models I_{n}$ and we have derived a contradiction.

\medskip

\noindent
Case 2. $L \subseteq Log(\mathcal{F}_{\text{S5}})$. Assume for a proof by contradiction that $\chi_{n}$ is equivalent to $I_{n}$ in $L$ and of size $<2^{n}$. 

We construct a model $\M_{2,n}$ based on $\F_{2,n}=(W_{2,n},R_{2,n})$ as follows.
Define $W_{2,n}= M_{2,n}\cup N_{2,n}$ with $M_{2,n}= \{a_{0},\ldots,a_{2^{n}-1}\}$ and $N_{2,n}=\{b_{0},\ldots,b_{2^n-1}\}$ and let $R_{2,n}=W_{2,n}\times W_{2,n}$.
Then $\F_{2,n}$ validates $L$. Let the valuation of $\M_{2,n}$ be defined in the same way as the valuation of $\M_{1,n}$ in Case~1.

We have $\M_{2,n},a_{i}\models I_{n}$ for all $i<2^{n}$. Hence we have $\M_{2,n},a_{i}\models\chi_{n}$.
Pick for every $\xi\in \text{sub}(\chi_{n})$ satisfied in some world in $N_{2,n}$ a world $b_{\xi}$ in $N_{2,n}$ satisfying $\xi$ in $\M_{2,n}$. Also pick for every $\xi\in \text{sub}(\chi_{n})$ satisfied in some world in $M_{2,n}$ a world $a_{\xi}$ in $M_{2,n}$ satisfying $\xi$ in $\M_{2,n}$. Let $\M_{i}$ denote the restriction of $\M_{2,n}$ to 
\[
\{a_{i}\}\cup \{ b_{\xi}\mid b_{\xi} \in \text{sub}(\chi_{n}) \text{ picked}\} \cup \{ a_{\xi}\mid a_{\xi} \in \text{sub}(\chi_{n}) \text{ picked}\}.
\]
Now the proof continues as in Case 1:
Clearly each $\M_{i},c$ with $c$ in $\M_{i}$ satisfies the same type $t\in T_{n}$ as $\M_{2,n},c$ and it is easy to see that $\M_{i},a_{i}\models \chi_{n}$ for all $i<2^{n}$. Hence $\M_{i},a_{i}\models I_{n}$ for $i<2^{n}$. However, there is a type in $T_{n}$ that is not satisfied in any $b_{\xi}$ since the number of $b_{\xi}$ in $\M_{i}$ is $<2^{n}$. Hence there is an $a_{i}$ with $\M_{i},a_{i}\not\models I_{n}$ and we have derived a contradiction.
The remaining cases are considered in the appendix.

\medskip

The logic Alt$_{1}$ can be used to show that there are non-tabular normal modal logics for which an exponential lower bound for the size of uniform interpolants would imply P $\not=$ NP. The proof is again by reduction to uniform interpolants for propositional logic. Observe that Alt$_{1}$ has the poly-size model property but does not have the exponential growth property.

\begin{restatable}{theorem}{thmaltalternative}
\label{thm:alt1_alternative}
Let $\sigma$ be a signature, and $\widehat{\sigma}=\sigma \times \mathbb{N}$. There are poly-time translations
\[\tr_\text{alt}: \text{ML}(\sigma) \to \text{PL}(\widehat{\sigma})\quad \text{ and } \quad \rt_\text{alt}: \text{PL}(\widehat{\sigma})\to \text{ML}(\sigma)\]
such that for every modal formula $\phi$, every $\tau \subseteq \sigma$ and every uniform $\tau\times \mathbb{N}$-interpolant $\psi$ for $\tr_\text{alt}(\phi)$ in propositional logic, $rt_\text{alt}(\psi)$ is a uniform $\text{Alt}_1(\tau)$-interpolant for $\phi$.

Here, if $\phi\in \text{ML}
(\sigma)$ has modal depth $n$, then $\tr_\text{alt}(\varphi)\in \text{PL}(\sigma\times \{0,\ldots,n\})$.
\end{restatable}
\begin{proof}
We adapt the translation developed in Section~\ref{sec:toprop}. However, naively applying the method from Section~\ref{sec:toprop}, would result in an exponentially sized interpolant. We therefore define a simpler translation that relies on the fact that one can work with finite frames for Alt$_{1}$ without cycles and that on such frames bisimulations are trivial: between any two rooted models there is at most one bisimulation and, if it exists, it is an isomorphism. 

Let $\phi\in \ML(\sigma)$ have modal depth $n$ and assume $\tau\subseteq \sigma$. The translation $\tr_{\text{alt}}(\varphi)$ will only be concerned
with the frames $\mathcal{F}_n=\{\F_0,\cdots, \F_n\}$ where $\F_k=(W_k,R_k)$ is given by $W_k = \{k,\cdots, 0\}$, $R_k =\{(i,i-1)\mid 0< i\leq  k\}$ and the root of $\F_i$ is the world $i$ (so here we do not follow the convention that the root is labeled 0). In other words, $\mathcal{F}_n$ is the set of linear chains of length up to $n$, with the worlds counting down from $k$. We adapt the technique developed in Section~\ref{sec:toprop} to find a strongest Alt$_{1}(\tau)$-implicate which, because Alt$_1$ has CIP, will also be a uniform Alt$_{1}(\tau)$-interpolant. 

Let $\F_i\in \mathcal{F}_n$, and let $\M$ be any model based on $\F_i$. We define the associated propositional model $v_\M$ over $\sigma\times \{0,\cdots, i\}$ by taking $v_\M(p_j)=1$ iff $\M,j\models p$. Additionally, we recursively define a set of translations $\tr_i: \text{ML}(\sigma)\to \text{PL}(\widehat{\sigma})$, for $i\in \mathbb{N}$, by 
\begin{align*}
    \tr_{i}(p)={} & p_{i},\\
    \tr_{i}(\neg \phi)={} &\neg \tr_{i}(\phi),\\
    \tr_{i}(\phi_1\vee\phi_2) = {} &\tr_{i}(\phi_1)\vee \tr_{i}(\phi_2),\\
    \tr_{i}(\square \phi)={} &\left\{\begin{array}{ll}
    \top & \text{ if }i=0\\
    \tr_{i-1}(\phi)&\text{ if }i>0\end{array}\right.
\end{align*}
and a reverse translation $\rt_i: \text{PL}(\sigma \times \{0,\cdots, i\})\rightarrow \text{ML}(\sigma)$ by
\[\rt_i(\psi) = \psi[p_j\mapsto \lozenge^{i-j} p].\]

It is straightforward to verify that for every $\chi\in \ML(\sigma)$, $v_\M\models \tr_i(\chi)$ if, and only if, $\M,i\models \chi$, and that for every propositional formula $\psi\in \text{PL}(\sigma\times \{0,\cdots,i\})$, $v_\M\models \psi$ if, and only if, $\M,i\models \rt_i(\psi)$.

We next define a strongest Alt$_{1}(\tau)$-implicate for $\phi$ using uniform propositional $\tau\times\{0,\cdots, i\}$-interpolants for $\tr_i(\phi)$, for $0\leq i \leq n$. Let, for each $0\leq i \leq n$, $\psi_i$ be a uniform propositional $\tau\times\{0,\cdots, i\}$-interpolant for $\tr_i(\phi)$. Consider the modal formula
\[\psi = \bigwedge_{0\leq i \leq n-1}(\lozenge^i\square\bot\rightarrow \psi_i[p_j\mapsto \lozenge^{i-j}p])\wedge (\lozenge^n\top\rightarrow \psi_n[p_j\mapsto \lozenge^{n-j}p])\]
which uses propositional variables in $\tau$ only. It is straightforward to show that $\psi$ is a strongest Alt$_{1}(\tau)$-implicate for $\phi$, as required. 

We have used multiple translations $\tr_i$ and $rt_{i}$ and also multiple uniform propositional interpolants, as was previously done for tabular logics determined by sets of finite frames in Section~\ref{subsec:there_and_back}. A single translation $\tr_{\text{alt}}$ and single converse translation $rt_{\text{alt}}$ are obtained in the same way as in Section~\ref{subsec:single_to_multiple}.
\end{proof}
We note that it is also open whether the dichotomy above can be extended to K4 since the pre-tabular normal logics containing K4 are significantly more complex than the pre-tabular normal modal logics containing S4 or GL. For instance, there are pre-tabular 
normal modal logics containing K4 without the poly-size model property~\cite[Chapter~12]{DBLP:books/daglib/0030819}.
\section{Conclusion}
\label{sec:conclusion}

We have investigated size bounds for Craig interpolants, uniform interpolants, and strongest implicates, for various families of normal modal logics, and, more generally, quasi-normal modal logics.

For \emph{tabular} quasi-normal modal logics, our results show that size upper bounds for strongest implicates are essentially the same as in the case of propositional logic  (modulo polynomial translations). This is, in some sense, 
the best one can hope for: a polynomial upper bound would
imply that $\text{NP} \subseteq \text{P}/\text{poly}$, 
a longstanding open problem in complexity theory, whereas
a super-polynomial lower bound would imply that $\text{NP} \not\subseteq \text{P}/\text{poly}$.
We obtained analogous results regarding Craig interpolants and uniform interpolants (for tabular quasi-normal modal logics that have CIP or, equivalently, UIP).

Complementing these results for tabular logics, we established  exponential lower bounds on the size of strongest implicates, uniform interpolants, and
Craig interpolants, which apply to all
\emph{non-tabular} normal logics that are contained in or contain S4 or GL (or, more generally, that have a non-tabular intersection with S4 or GL). This covers most non-tabular normal modal logics in practice,
and matches known single-exponential upper bounds on the size of Craig interpolants for common modal
logics such as K, K4, GL, S5, and Grz, as well as known single-exponential upper bounds on the size of uniform interpolants for S5 and K. The problem of finding tight bounds on the size of uniform interpolants for GL and Grz remains open.

The main problem left open is whether one can establish a dichotomy of the following form for all (quasi-)normal modal logics: either computing strongest implicates is poly-time reducible to computing uniform interpolants in classical propositional logic or strongest implicates are of at least exponential size, and similarly for Craig interpolants for modal logics having CIP. The logic Alt$_{1}$ shows that to prove this, one has to replace tabularity by some weaker condition. For normal modal logics containing K4, however, we conjecture that tabularity can still be used to show a dichotomy. Another intriguing problem is the size of Craig interpolants (when they exist) for tabular modal logics without CIP. We have seen that our current reduction does not work in this case, but there could still be a completely different approach. On the other hand, it is known that Craig interpolants often become much harder to compute for logics without CIP (and uniform interpolants for logics without UIP)~\cite{DBLP:conf/lics/JungW21,DBLP:conf/kr/JungLPW21,jung2025computation}. 

As discussed earlier, these results all pertain to formula-size as measured based on a dag representation.
Interestingly, it is not difficult to show that our reduction of strongest implicates in tabular quasi-normal modal logics to uniform interpolants in propositional logic (and, similarly, of Craig interpolants for tabular logics with CIP) provides `propositional sized' implicates also if a standard tree-representation is used. In fact, while the tree-size of the `forward-translation' $tr_{L}(\varphi)$ of a formula $\varphi$ can be exponential, the `backward-translation' $rt_{L}(\varphi)$ is always of polynomial tree-size. The latter suffices to obtain strongest implicates of polynomial tree-size in the size of the propositional uniform interpolant and 
$tr_{L}(\varphi)$ can always be represented, using additional propositional variables, as a formula of polynomial tree-size. 
This observation regarding $rt_{L}$ can be used to lift further properties from propositional logic to tabular modal logics. For instance, assume that every propositional formula of dag-size $n$ is equivalent to a propositional formula $\psi$ of tree-size $f(n)$. Then, in tabular modal logics,
every modal formula of dag-size $n$ is equivalent to a modal formula of tree-size $poly(f(poly(n)))$. 

Note that, for the modal logic K (which has UIP), 
it is known that the tree-size of 
uniform interpolants (and hence also strongest $\sigma$-implicates and Craig interpolants) 
can be bounded by an exponential function
in the tree-size of  input formula(s). 
On the other hand, it remains an open problem whether the exponential upper bounds for Craig interpolants in K4, S4, GL, and Grz still hold if one uses a tree representation of interpolants.

\bibliographystyle{plainurl}
\bibliography{main}

\newpage

\appendix

\section{Proofs for Section~\ref{sec:toprop}}

\lemmacoveramount*
\begin{proof}
Let $\M=(\F,V)$ be a model based on $\F$. We can obtain a $\sigma$-cover for $\M$ by using a refinement procedure. We begin with $\Psi_0 = \{\top\}$, which is a $\sigma$-EME. Then, we obtain $\Psi_{m+1}$ from $\Psi_m$ as follows. If, for some $\psi\in \Psi_m$ and $p\in \sigma$, both $\psi\wedge p$ and $\psi \wedge \neg p$ are satisfied in $\M$, we let $\Psi_{m+1}=\Psi_m\setminus \{\psi\}\cup \{\psi\wedge p, \psi\wedge \neg p\}$. Otherwise, we let $\Psi_{m+1}=\Psi_m$. This process will terminate after at most $|W|-1$ refinements, call the result $\Phi$. This $\Phi$ is by construction a $\sigma$-EME and any two elements of $\M$ that satisfy the same $\psi$ are propositionally indistinguishable over $\sigma$, since otherwise further refinement would have been possible. Hence $\Phi$ is a $\sigma$-cover of $\M$.

Note that not every $\sigma$-cover is of the form generated by this procedure, but every $\M$ has at least one such $\sigma$-cover. Furthermore,  every $\psi\in \Phi$ is a conjunction of at most $|W|$ literals, and hence $|\psi|\leq O(|W|)$. 

To find a bound on the number of $\sigma$-covers we need, it is convenient to represent them as decision trees that determine which element of $\Phi$ a particular world satisfies, based on which formulas are true or false in that world. Formally, a cover is then represented as a labeled tree $(\mathit{vert},\mathit{edge},\mathit{label})$ where $\mathit{edge}\subseteq \mathit{vert}\times \mathit{vert}$ and $\mathit{label}$ is a function that assigns each vertex $x\in \mathit{vert}$ a formula $\mathit{label}(x)$. 

Corresponding with $\Psi_0$, we take $T_0 = (\mathit{vert}_0,\mathit{edge}_0,\mathit{label}_0)$ where $\mathit{vert}=\{x_0\}$, $\mathit{edge}_0=\emptyset$ and $\mathit{label}_0(x_0)=\top$. Then, if from $\Psi_m$ to $\Psi_{m+1}$ we split $\psi$ into $\psi\wedge p$ and $\psi\wedge\neg p$, then we take 
\begin{itemize}
	\item $\mathit{vert}_{m+1}=\mathit{vert}_m\cup \{x_m,y_m\}$, 
	\item $\mathit{edge}_{m+1}=\mathit{edge}_m\cup \{(x,x_m), (x,y_m)\}$, where $x$ is the leaf such that $\mathit{label}_m(x)=\psi$, and
	\item $\mathit{label}_{m+1}(x)= \left\{\begin{array}{ll}\psi\wedge p & \text{ if } x=x_m\\\psi\wedge \neg p & \text{ if } x=y_m\\ \mathit{label}_m(x) & \text{ otherwise}\end{array}\right.$
\end{itemize}
and if $\Psi_m= \Psi_{m+1}$ then we take $T_m = T_{m+1}$. Call the final tree $T=(\mathit{vert},\mathit{edges},\mathit{label})$.

The unlabeled tree $\overline{T}=(\mathit{vert},\mathit{edges})$ is a sub-tree of the binary tree of depth $|W|$, and has at most $|W|$ leaves. Each such unlabeled tree can therefore be identified by choosing at most $N$ leaves from among the $2^{|W|}$ nodes of the binary tree. So there are at most $\binom{2^{|W|}}{|W|}\leq \left(2^{|W|}\right)^{|W|}=2^{|W|^2}$ unlabeled trees. Any labeling is determined by the atoms that were used, in each non-leaf vertex, to differentiate between the two successors of that vertex. There are at most $|W|-1$ such non-leaf vertices, so there are at most $|\sigma|^{|W|-1}$ labelings for a given unlabeled tree. There are, therefore, at most $2^{|W|^2}\cdot |\sigma|^{|W|-1}$ different $\sigma$-covers of this type. Since every model has such a $\sigma$-cover, the same number bounds the amount of $\sigma$-covers that we need.

The time bound follows because we can construct $\Gamma_\F$ by iterating over all labeled trees.
\end{proof}

\propsinglemodelbisim*
\begin{proof}
For $(1)\Rightarrow (2)$, suppose that $Z:(\M,0)\bisim_\sigma (\M',0')$. As $Z$ is a bisimulation between the roots of $\M$ and $\M'$, it must be a total relation. Hence for every $w'\in W$ there is some $w\in W$ such that $wZw'$. Because $Z$ is a bisimulation, this implies that 
$V(w)\cap \sigma =V'(w')\cap \sigma$. This implies that $\Phi$ is a $\sigma$-cover of $\M'$ as well as of $\M$, so $\M'$ has a $\Phi$-abstraction. Let $\A'=(\F',f')$ be the $\Phi$-abstraction of $\M'$.

Because $Z$ is a bisimulation between $\M,0$ and $\M',0$, it satisfies {\bf back} and {\bf forth}. In order to show that it is an abstract bisimulation between $\A,0$ and $\A',0$ it therefore suffices to show that $Z$ satisfies {\bf abstract atoms}.

Suppose, therefore, that $w Z w'$. Because $Z$ is a $\sigma$-bisimulation, we have $\M,w\models p$ iff $\M',w'\models p$ for all $p\in \sigma$. As each $\phi\in \Phi$ is a propositional formula over $\sigma$, it follows that $\M,w\models \phi$ iff $\M',w'\models \phi$.

We therefore have $f(w)=\phi \Leftrightarrow \M,w\models \phi \Leftrightarrow \M',w'\models \phi \Leftrightarrow f'(w')=\phi$. This holds for all $\phi\in \Phi$, so {\bf abstract atoms} is satisfied, as was to be shown.

For $(2)\Rightarrow (1)$, suppose that $Z: (\A,0)\bisim_\Phi (\A',0)$. As $Z$ is an abstract bisimulation between the roots of $\A$ and $\A'$, it must be a total relation.
For any $w'\in W'$, fix a single $w_{w'}\in W$ such that $w_{w'}Z w'$, and let $\M'=(\F',V')$ be the model such that $V'(w')=V(w_{w'})$. We will show that $Z$ is a total $\sigma$-bisimulation between $\M$ and $\M'$, that $\Phi$ is a $\sigma$-cover of $\M'$ and that $\A'$ is the $\Phi$-abstraction of $\M'$.

Because $Z$ is an abstract bisimulation, it satisfies {\bf back} and {\bf forth}. To show that it is a $\sigma$-bisimulation, it therefore suffices to show that it satisfies {\bf atoms}. Take any $w\in W, w'\in W'$ such that $w Z w'$. 

Then both $wZ w'$ and $w_{w'}Z w'$. So $f(w)=f'(w')=f(w_{w'})$, because $Z$ is an abstract bisimulation. Since $\Phi$ is a $\sigma$-cover of $\M$ it follows from $f(w)=f(w_{w'})$ that $V_\sigma(w)=V_\sigma(w_{w'})$. Furthermore, by construction, $V(w_{w'})=V'(w')$. It follows that $V_\sigma(w)=V'_\sigma(w')$, so {\bf atoms} is satisfied, as was to be shown.

Now, to show that $\Phi$ is a $\sigma$-cover of $\M'$. 
Take any $w_1',w_2'$ such that $\M',w_1'\models \phi$ and $\M',w_2'\models \phi$, for some $\phi\in \Phi$. Then $\M,w_{w_1'}\models \phi$ and $\M,w_{w_2'}\models \phi$, so because $\Phi$ is a $\sigma$-cover of $\M$ we have $\M,w_{w_1'}\equiv_{\text{PL}(\sigma)}\M,w_{w_2'}$. By construction we also have $\M,w_{w_1'}\equiv_{\text{PL}(\sigma)}\M',w_1'$ and $\M,w_{w_2'}\equiv_{\text{PL}(\sigma)}\M',w_2'$, so we obtain $\M',w_1'\equiv_{\text{PL}(\sigma)}\M',w_2'$. So $\Phi$ is a $\sigma$-cover of $\M'$.

Finally, take any $w'\in W'$ and let $\phi = f'(w')$. Because $w_{w'}Z w'$ and $Z$ is both an abstract bisimulation between $\A$ and $\A'$ and a $\sigma$-bisimulation between $\M$ and $\M'$, we have $f(w_{w'})=f'(w')=\phi$, and $\M,w_{w'}\models \phi$ iff $\M',w'\models \phi$. Also, because $\A$ is the $\Phi$-abstraction of $\M$, we have $\M,w_{w'}\models f(w_{w'})$. Taken together, this shows that $\M',w'\models f'(w')$, which completes the proof that $\A'$ is the $\Phi$-abstraction of $\M'$.
\end{proof}

\propidentifiersize*
\begin{proof}
Let $\A=(\F,f)$ with $\F=(W,R)$.
Recall that, for $w\in W$, $[w]=\{w'\mid \A,w \sim_{\Phi} \A,w'\}$. We begin by constructing, for every $w\in W$, a characteristic formula $\psi_{[w]} \in \text{ML}(\Phi)$ such that $\A,w'\models \psi_{[w]}$ if, and only if, $w'\in [w]$. We do this through the standard partition refinement procedure.
For every $\phi\in \Phi$, we let $W_\phi = \{w\in W\mid f(w)=\phi\}$, and we take $\mathcal{I}_0 = \{(W_\phi,\phi)\mid \phi\in \Phi\}$. Then, from $\mathcal{I}_m$ we obtain $\mathcal{I}_{m+1}$ by taking any two $(J,\psi), (J',\psi')\in \mathcal{I}_m$ and checking whether there are $j_1,j_2\in J$ such that there exists $j_1'\in J'$ with $(j_1,j_1')\in R$ but $(j_2,j_2')\not \in R$ for all $j_2'\in J'$.
If so, then in $\mathcal{I}_{m+1}$ we replace the pair $(J,\psi)$ by the two pairs
\[
(\{j\in J\mid \exists j'\in J':(j,j')\in R\}, \psi\wedge\lozenge\psi') \] 
and  
\[((\{j\in J\mid \forall j'\in J':(j,j')\not\in R\}, \psi\wedge\neg\lozenge\psi').\] 
If no such $J, J', j_1, j_2$ and $j_1'$ exist we stop and set $\mathcal{I}=\mathcal{I}_m$. 

Because we obtained $\mathcal{I}=\{(I_1,\psi_1),\cdots,(I_k,\psi_k)\}$ through a partition refinement procedure, we know that $I_1, \cdots, I_k$ form a partition of $W$, and that $\A,w\sim_{\Phi}\A,w'$ if and only if there is some $I_i$ such that $w,w'\in I_i$. This implies that if $w\in I_i$, then $[w]=I_i$. 

For any $w\in W$, let us therefore write $\psi_{[w]}$ for the unique formula such that there is a $J$ such that $(J,\psi_{[w]})\in \mathcal{I}$ and $w\in J$. By the construction of $\mathcal{I}$, we have $\A,w'\models \psi_{[w]}$ if and only if $w'\in J=[w]$. So $\psi_{[w]}$ is indeed the characteristic formula that we were looking for.

A straightforward induction on $n$ shows that,
for each $(J,\psi)\in \mathcal{I}_n$ 
($n\geq 0$), it holds that $|\psi|\leq O(2^n)\cdot \max_{\phi\in\Phi}|\phi|$. 
Furthermore, since in each iteration the partition is refined and each $\mathcal{I}_n$ can contain at most $|\F|\leq N$ elements,
the above process must stabilize after at most $m< N$ iterations. We can therefore conclude that
each $\psi_i$, for  $i\leq k$, has size at most
$O(2^N)\cdot  \max_{\phi\in\Phi}|\phi|$.

Now, let 
\begin{align*}
\charf(\mathcal{I},\F) = {} & \Big( \Boxmulti{N} \bigvee_{(J,\psi)\in \mathcal{I}}\psi \Big) \wedge \\&\Big(\bigwedge_{(J,\psi)\in \mathcal{I}}\bigwedge_{(J',\psi')\in \mathcal{I}}\Boxmulti{N} (\psi\rightarrow \pm \lozenge \psi')\Big),
\end{align*}
where $\pm \lozenge \psi'$ is $\lozenge \psi'$ if there are $j\in J, j'\in J'$ such that $(j,j')\in R$ and $\neg \lozenge \psi'$ otherwise, and take
\begin{align*}
\delta_\A = {} & \charf (\mathcal{I},\F)\wedge  \psi_{[0]}.
\end{align*}

It is not difficult to see that 
$|\delta_\A|\leq O(2^N)\cdot \max_{\phi\in\Phi}|\phi|$ and $\delta_\A$ can be computed
in time $O(2^N)\cdot\Sigma_{\phi\in\Phi}|\phi|$.

It remains only to show that for every rooted abstract $\Phi$-model $\A',0$ of size at most $N$, we have $\A',0\models \delta_\A$ iff $\A',0\bisim_\Phi \A,0$.

$(\Rightarrow)$ Suppose that $\A',0\models \delta_\A$. Let $Z \subseteq W\times W'$ be the relation such that $wZ w'$ iff $\A',w'\models \psi_{[w]}$. We will show that $Z$ is an abstract $\Phi$-bisimulation between $\A$ and $\A'$ and that $0 Z 0$.

That $0Z 0$ is the easy part: one of the conjuncts of $\delta_\A$ is $\psi_{[0]}$, so from $\A',0\models \delta$ it follows that, in particular, $\A',0\models \psi_{[0]}$ and therefore $0Z 0$. We continue to show that $Z$ is an abstract $\Phi$-bisimulation.

For {\bf abstract atoms}, suppose that $wZ w'$. We have $\A,w\models \psi_{[w]}$ by the construction of $\psi_{[w]}$. By the construction of $Z$ it follows from $wZ w'$ that we also have $\A',w'\models \psi_{[w]}$. One conjunct of $\psi_{[w]}$ is some $\phi\in \Phi$, so we have $\A,w\models \phi$ and $\A',w'\models \phi$, so $f(w)=f'(w')=\phi$, as required for {\bf abstract atoms} to hold.

For {\bf forth}, suppose that $w_1Z w_1'$ and $(w_1,w_2)\in R$. Because any world can be reached from 0 in at most $N$ steps, it follows from $\A',0\models \delta_\A$ that $\A',w_1'\models \psi_{[w_1]}\rightarrow \pm \lozenge \psi_{[w_2]}$. Furthermore, because $(w_1,w_2)\in R$ the $\pm$ is positive, so $\A',w_1'\models \psi_{[w_1]}\rightarrow \lozenge \psi_{[w_2]}$.

From $w_1 Z w_1'$ it follows that $\A',w_1'\models \psi_{[w_1]}$, which implies that $\A',w_1'\models \lozenge \psi_{[w_2]}$. Hence there is some $w_2'\in W'$ such that $(w_1',w_2')\in R'$ and $\A',w_2'\models \psi_{[w_2]}$. This implies that $w_2Z w_2'$. We now have $(w_1',w_2')\in R'$ and $w_2 Z w_2'$, so {\bf forth} is satisfied.

For {\bf back}, suppose that $w_1Z w_1'$ and $(w_1',w_2')\in R'$. Because $\M',0\models \charf(\mathcal{I},\F)$ and $w_2'$ can be reached from $0$ in at most $N$ steps, we have $\M',w_2'\models \psi$ for some $(J,\psi)\in \mathcal{I}$. This implies that $\M',w_1'\models \lozenge \psi$ and therefore $\M',w_1'\not\models \psi_{[w_1]}\rightarrow \neg \lozenge \psi$. We must therefore have $\pm \lozenge \psi = \lozenge \psi$, so there is some $w_2\in J$ such that $\psi_{[w_2]}=\psi$ and $(w_1,w_2)\in R$. This implies that $(w_1,w_2)\in R$ and $w_2 Z w_2'$, so {\bf back} is satisfied. This completes the left-to-right direction of the proof that $\delta_\A$ is the abstract class identifier for $[\A]$.

$(\Leftarrow)$ Suppose that $Z:\A,0\bisim_\Phi \A',0$. To show is that $\A',0\models \delta_\A$.

In order to do so, first note that because $Z$ is an abstract $\Phi$-bisimulation, and therefore satisfies {\bf abstract atoms}, we have for every $(W_\phi,\phi)\in \mathcal{I}_0$, every $w\in W_\phi$ and every $w Z w'$ that $\A',w'\models \phi$.

Because of the way the $\mathcal{I}_m$ are created through a partition refinement procedure, it follows from {\bf back} and {\bf forth} that for every $\mathcal{I}_m$, every $(J,\psi)\in \mathcal{I}_m$, every $w\in J$ and every $w Z w'$ that $\A',w'\models \psi$.

It immediately follows that $\A',0\models \psi_{[0]}$. Furthermore, because $Z$ is an abstract bisimulation that relates $\A,0$ and $\A',0$, it follows from the fact that all worlds are reachable from $0$ that $Z$ is a total relation. So for every $w'\in W'$ there is a $w\in W$ such that $wZ w'$. There is also some $(J,\psi)\in \mathcal{I}$ such that $w\in J$. It follows that $\A',w'\models \psi$, and therefore $\A',0\models \bigwedge_{0\leq l \leq N}\square^l\bigvee_{(J,\psi)\in \mathcal{I}}\psi$.

Now, take any $(J,\psi), (J',\psi')\in \mathcal{I}$. We need to show that, for any $w_1'\in W'$, we have $\A',w'\models \psi\rightarrow \pm \lozenge \psi'$. So suppose that $\A',w'\models \psi$. Then $wZ w'$ for some $w\in J$. If $\pm \lozenge \psi'=\lozenge \psi'$ then there is some $w_2\in J'$ such that $(w_1,w_2)\in R$. By {\bf forth} there is then some $w_2'\in W'$ such that $w_2 Z w_2'$ and $(w_1',w_2')\in R'$. Because $w_2\in J'$ and $w_2 Z w_2'$ we have $\A',w_2'\models \psi'$, and hence $\A',w_1'\models \lozenge \psi'$, so $\A',w_1'\models \psi\rightarrow \pm\lozenge \psi'$.

If $\pm \lozenge \psi'=\neg\lozenge \psi'$ then for all $w_2\in J'$ we have $(w_1,w_2)\not \in R$. By {\bf back} it follows that for all $w_2'\in W'$ such that $w_2 Z w_2'$ we have $(w_1',w_2')\not\in R'$. This implies that for every $w_3'$ such that $(w_1',w_3')\in R'$ there is some $w_3\not \in J'$ such that $w_3 Z w_3'$. Therefore, $\A',w_3'\not\models \psi'$. As this holds for all such $w_3'$, we have $\A',w_1'\models \neg \lozenge \psi'$ and therefore $\A',w_1'\models \psi\rightarrow \pm \lozenge \psi'$.

We have now shown that $\A',w_1'\models \psi\rightarrow \pm \lozenge \psi'$ for all $w_1'\in W'$ and all $(J,\psi), (J',\psi')\in \mathcal{I}$. This implies that

\[\A',0\models \bigwedge_{(J,\psi)\in\mathcal{I}}\bigwedge_{(J',\psi')\in\mathcal{I}}\Boxmulti{N}(\psi\rightarrow \pm \lozenge \psi').\]
Together with the previously shown $\A',0\models \Boxmulti{N}\bigvee_{(J,\psi)\in \mathcal{I}}\psi$ and $\A',0\models \psi_{[0]}$ this shows that $\A',0\models \delta$, which completes the right-to-left direction.
\end{proof}

\thmmain*

\begin{proof}
We extend the translation $tr_{\F}$ to a translation $tr_{L}$ for any tabular quasi-normal logics $L$. Fix $L=\text{Log}(\mathcal{F}_{L})$ for a finite set $\mathcal{F}_{L}$ of finite rooted frames. We assume that for $(\F,w),(\F',w')\in \mathcal{F}_{L}$ with $(\F,w)\not=(\F',w')$ the sets of worlds in $\F$ and $\F'$ are mutually disjoint (and so do not use $0$ to denote the roots of frames in $\mathcal{F}_{L}$).
Take atoms $r_{\F,w}$, for $(\F,w)\in \mathcal{F}_{L}$. They are used to identify the frame $(\F,w)\in \mathcal{F}_{L}$ in which we evaluate a modal formula. For any
signature $\sigma$, let $\sigma_{L}= \bigcup_{(\F,w)\in \mathcal{F}_{L}}\sigma_{\F,w}$. The signatures  $\sigma_{\F,w}$ are assumed to be mutually disjoint and also disjoint from $\{r_{\F,w}\mid (\F,w)\in \mathcal{F}_{L}\}$. Let $\widehat{\sigma}= \sigma_{L}\cup \{r_{\F,w}\mid (\F,w) \in \mathcal{F}_{L}\}$.

Next let for any modal formula $\varphi$,
\[
tr_{L}(\varphi)= \text{Unique}(\mathcal{F}_{L})
\wedge \bigwedge_{(\F,w) \in \mathcal{F}_{L}} (r_{\F,w}\rightarrow tr_{\F,w}(\varphi)).
\]
where 
\[\text{Unique}(\mathcal{F}_{L})=(\bigvee_{(\F,w)\in \mathcal{F}_{L}} (r_{\F,w} \wedge \bigwedge_{(\mathfrak{G},v)\in \mathcal{F}_{L}\setminus \{(\F,w)\}} \neg r_{\mathfrak{G},v}))
\]
Note that if $\varphi\in \ML(\sigma)$, then $tr_{L}(\varphi) \in \PL(\widehat{\sigma})$. 
For any model $\M$ based on a frame $(\F,w)\in \mathcal{F}_{L}$ we obtain a propositional model $v_{\M}$ defined by setting 
\begin{itemize}
    \item $v_{\M}(r_{\F,w})=1$ and $v_{\M}(r_{\mathfrak{G},v})=0$ for $(\mathfrak{G},v)\in \mathcal{F}_{L}\setminus \{(\F,w)\}$;  
    \item $v_{\M}(p_{\F,v})=1$ iff $\M,v\models p$, for all $v$ in $\F$; 
    \item $v_{\M}(p_{\mathfrak{G},v'})\in \{0,1\}$ arbitrary 
    for all worlds $v'$ in any $(\mathfrak{G},v)\in \mathcal{F}_{L}\setminus\{(\F,w)\}$.
\end{itemize}
\begin{claim}\label{claim:1}
    Let $(\F,w)\in \mathcal{F}_{L}$ and $\M$ be a model based on $\F$. Then
    \begin{enumerate}
        \item for every modal formula $\varphi$, we have 
    $\M,0\models \varphi$ iff $v_{\M}\models tr_{L}(\varphi)$.
        \item For every modal implication $\varphi \rightarrow \psi$, we have $\varphi \rightarrow \psi\in L$ iff $tr_{L}(\varphi) \rightarrow tr_{L}(\psi)$ is a tautology.
    \end{enumerate}
\end{claim}
The proof is straightforward. 
The following claims imply that we can transform any propositional uniform $\widehat{\sigma}$-interpolant of $tr_{L}(\varphi)$ in polynomial
time into an equivalent one of the form $\text{Unique}(\mathcal{F}_{L}) \wedge  \bigwedge_{(\F,w) \in \mathcal{F}_{L}} (r_{(\F,w)}\rightarrow \xi_{\F,w})$ with $\xi_{\F,w}$ a propositional uniform $\sigma_{\F,w}$-interpolant of $tr_{\F,w}(\varphi)$. 
\begin{claim}\label{claim:transforminter}
    Let $\varphi$ be a modal formula and $\sigma\subseteq \text{sig}(\varphi)$. Let $\xi$ be any propositional uniform $\widehat{\sigma}$-interpolant of $tr_{L}(\varphi)$ and let $\xi_{\F,w}$ be any propositional uniform $\sigma_{\F,w}$-interpolant of $tr_{\F,w}(\varphi)$, for $(\F,w)\in \mathcal{F}_{L}$. Then $\xi$ and $\text{Unique}(\mathcal{F}_{L}) \wedge  \bigwedge_{(\F,w) \in \mathcal{F}_{L}} (r_{\F,w}\rightarrow \xi_{\F,w})$ are logically equivalent.    
\end{claim}
To show Claim~\ref{claim:transforminter}, it suffices to show that \[\text{Unique}(\mathcal{F}_{L}) \wedge \bigwedge_{(\F,w) \in \mathcal{F}_{L}} (r_{\F,w}\rightarrow \xi_{\F,w})\] is a uniform $\widehat{\sigma}$-interpolant of $tr_{L}(\varphi)$ (since uniform interpolants are unique up to logical equivalence). We first show that 
\[
tr_{L}(\varphi) \rightarrow (\text{Unique}(\mathcal{F}_{L}) \wedge  \bigwedge_{(\F,w) \in \mathcal{F}_{L}} (r_{\F,w}\rightarrow \xi_{\F,w}))
\]
is a tautology. Consider a model $v$ satisfying $tr_{L}(\varphi)$.
Then we have a unique $(\F,w)\in \mathcal{F}_{L}$ with $v\models r_{\F,w}$. We also have $v\models tr_{\F,w}(\varphi)$. But then
also $v\models \text{Unique}(\mathcal{F}_{L}) \wedge  \bigwedge_{(\F,w) \in \mathcal{F}_{L}} (r_{\F,w}\rightarrow \xi_{\F,w})$. Next one can easily show that every model $v$ of 
$\text{Unique}(\mathcal{F}_{L}) \wedge  \bigwedge_{(\F,w) \in \mathcal{F}_{L}} (r_{\F,w}\rightarrow \xi_{\F,w})$
can be expanded to a model $v'$ of $tr_{L}(\varphi)$. The claim follows directly.

\medskip

Let $\varphi$ be a modal formula and $\sigma\subseteq \text{sig}(\varphi)$. Let $\xi$ be any propositional uniform $\widehat{\sigma}$-interpolant of $tr_{L}(\varphi)$. For $(\F,w)\in \mathcal{F}$, obtain $\xi^{\uparrow\F,w}$ from $\xi$ by
\begin{itemize}
    \item replacing $r_{\F,w}$ by $\top$,
    \item replacing all $r_{\mathfrak{G},v}$ with $(\mathfrak{G},v)\in \mathcal{F}_{L}\setminus\{(\F,w)\}$ by $\bot$, and
    \item replacing all $p_{\mathfrak{G},v}$ with $v$ not in $(\F,w)$ by $\bot$.
\end{itemize}
\begin{claim}\label{claim:transforminter2}
$\xi^{\uparrow\F,w}$ is a propositional uniform $\sigma_{\F,w}$-interpolant for $tr_{\F,w}(\varphi)$, for all $(\F,w)\in \mathcal{F}_{L}$.
\end{claim}
To show Claim~\ref{claim:transforminter2}, we first show that $tr_{\F,w}(\varphi) \rightarrow \xi^{\uparrow \F,w}$ is tautology. Let $v$ be  model of $tr_{\F,w}(\varphi)$. We obtain model of $tr_{L}(\varphi)$ by expanding $v$ to $v'$ with
\begin{itemize}
    \item $v'(r_{\F,w})=1$,
    \item $v'(r_{\mathfrak{G},v})=0$ for $(\mathfrak{G},v)\in \mathcal{F}_{L}\setminus\{(\F,w)\}$, and
    \item $v'(p_{\mathfrak{G},v})=0$ for $v$ not in $(\F,w)$.
\end{itemize}
But then $v'$ satisfies $\xi$ (since $\xi$ is a uniform $\widehat{\sigma}$-interpolant of $tr_{L}(\varphi)$). Hence, by definition, $v$ satisfies $\xi^{\uparrow \F,w}$. Next observe that every model of $\xi^{\uparrow \F,w}$ can be expanded to a model of $\xi$ is the obvious way. This model then satisfies $tr_{\F,w}(\varphi)$, by definition of the formulas. The claim follows directly.

\medskip

Define for any $\xi\in \PL(\widehat{\sigma})$, 
\[
rt_{L}(\xi) = \bigvee_{(\F,w) \in \mathcal{F}_{L}}rt_{(\F,w),L}(\xi^{\uparrow\F,w}).
\]
It follows from Theorem~\ref{thm:many-frame} and Claim~\ref{claim:transforminter2} that 
$rt_{L}$ is as required.
\end{proof}
\section{Proofs for Section~\ref{sec:CraigInt}}

\thmbsimCIP*

\begin{proof}
    We provide details not provided in the proof given in the paper.
    First, we show that Fact~1 implies the direction from right to left. Assume the right-hand side of the equivalence holds. We aim to show CIP. To apply Fact 1, assume $\varphi,\psi$ are given, $\sigma=\text{sig}(\varphi)\cap\text{sig}(\psi)$ and we have models $\M_{1},0\models \varphi$ and $\M_{2},0\models \psi$ with $\M_{1},0 \bisim_{\sigma} \M_{2},0$ such that $\M_{1}$ is based on $\F_{i}\in \mathcal{F}_{L}$ and $\M_{2}$ is based on $\F_{j}\in \mathcal{F}_{L}$. By (2), we have $i=j$ and we have an isomorphism $f:\F_{i} \rightarrow \F_{i}$ with $f(0)=0$ and $\M_{1},w\models p$ iff $\M_{2},f(w)\models p$ for all $w$ in $\F_{i}$ and $p\in \sigma$. But then we obtain a model $\M'$ based on $\F_{i}$ with $\M',0\models \varphi\wedge \psi$ by taking the $\text{sig}(\varphi)$-reduct of $\M_{1}$ and setting $\M',w\models p$ iff $\M_{2},f(w)\models p$ for all $p\in \text{sig}(\psi)\setminus\text{sig}(\varphi)$.

    We also show that $f_{1}$ and $f_{2}$ are invariant for $\sigma$-atoms in the sense that for all $w$ in $\F_{i}$ and $p\in \sigma$: $\M_{k},w\models p$ iff $\M',f_{k}(w)\models p$, for $k=1,2$.
    To prove invariance for $\M_{1}$, let $w$ in $\F_{i}$ and $p\in \sigma$. Observe that $\M_{1},w\models p$ iff $q_{w,1}\rightarrow p$ is a conjunct of $\delta_{\sigma}(\M_{1},0)$. We also have $\M',f_{1}(w)\models q_{w,1}$. Hence, if $\M_{1},w\models p$, then $\M',f_{1}(w)\models p$ since $\M',0\models \delta_{\sigma}(\M_{1},0)$. The same argument applies to $\neg p$ and so the equivalence for $\M_{1}$ follows. 
\end{proof}

\thmmainCIP*

\begin{proof}
The proof is similar to the proof of Theorem~\ref{thm:main} and uses Theorem~\ref{thm:many-frame-Craig}.
Fix $L=\text{Log}(\mathcal{F}_{L})$ with CIP and for a finite set $\mathcal{F}_{L}$ of finite rooted frames. We make the same assumptions as in the proof of Theorem~\ref{thm:main}.
For any signature $\sigma$, define $\hat{\sigma}$ as before.
Recall
\[
tr_{L}(\varphi)= \text{Unique}(\mathcal{F}_{L})
\wedge \bigwedge_{(\F,w) \in \mathcal{F}_{L}} (r_{\F,w}\rightarrow tr_{\F,w}(\varphi)).
\]
Now we show that the formulas $\xi^{\uparrow\F,w}$ introduced in the proof of 
Theorem~\ref{thm:main} for $\sigma=\text{sig}(\varphi)\cap \text{sig}(\psi)$ 
are interpolants for $tr_{\F,w}(\varphi)\rightarrow tr_{\F,w}(\psi)$ for $(\F,w)\in \mathcal{F}_{L}$ if $\xi$
is an interpolant for $tr_{L}(\varphi)\rightarrow tr_{L}(\psi)$.
\begin{claim}\label{claim:55}
Let $\varphi\rightarrow \psi$ be a modal implication and $\xi$ an interpolant for $tr_{L}(\varphi)\rightarrow tr_{L}(\psi)$ in $\sigma=\text{sig}(\varphi)\cap \text{sig}(\psi)$. Then $\xi^{\uparrow\F,w}$ is an interpolant for $tr_{\F,w}(\varphi)\rightarrow tr_{\F,w}(\psi)$, for every $(\F,w)\in \mathcal{F}_{L}$.
\end{claim}
The proof of Claim~\ref{claim:55} is similar to the proof of Claim~\ref{claim:transforminter2}.
One can first show in the same way that $tr_{\F,w}(\varphi) \rightarrow \xi^{\uparrow \F,w}$ is tautology, for all $(\F,w)\in \mathcal{F}_{L}$.
To show that $\xi^{\uparrow \F,w}\rightarrow tr_{\F,w}(\psi)$ is a tautology,
recall that every model of $\xi^{\uparrow \F,w}$ can be expanded to a model of $\xi$ in
the obvious way. This model then satisfies $tr_{L}(\psi)$ since $\xi$ is an interpolant.
But then it satisfies $tr_{\F,w}(\psi)$, as required. 

\medskip

It follows from Theorem~\ref{thm:many-frame-Craig} and Claim~\ref{claim:55} that 
$rt_{L}$ is as required.
\end{proof} 
\section{Proofs for Section~\ref{sec:lower}}
We first continue the proof of the second half of Theorem~\ref{thm:main-lower}.
\begin{theorem}
Let $L$ be any normal modal logic that is contained in a pre-tabular normal modal logic containing S4 or GL.
    There are modal formulas $(\phi_n)_{n=1,2,\ldots}$
 and 
 $(\psi_n)_{n=1,2,\ldots}$
 of size polynomial in $n$ 
 such that (i) $\phi_n\to\psi_n\in L$, (ii) a Craig interpolant for $\phi_n \to \psi_n$ exists, and (iii) every Craig interpolant for $\phi_n\to\psi_n$ in $L$ 
 has  size at least $2^n$.
 \end{theorem}
\begin{proof}
Define $\sigma_{n}$, $T_{n}$, $I_{n}$, $\varphi_{n}$, and $\psi_{n}$ as in the main paper.
We have already shown that for $L \subseteq Log(\mathcal{F}_{\text{Grz.3}})$ and $L \subseteq Log(\mathcal{F}_{\text{S5}})$ there are no $\chi_{n}$ equivalent to $I_{n}$ in $L$ of size $<2^{n}$.

\medskip

Case 3. $L \subseteq Log(\mathcal{F}_{1})$. 
Assume for a proof by contradiction that $\chi_{n}$ is equivalent to $I_{n}$ in $L$ and
of size $<2^{n}$. 

We construct a model $\M_{3,n}$ based on a frame $\F_{3,n}=(W_{3,n},R_{3,n})$ by adding to the
frame $\F_{2,n}$ defined in Case~2 a world $d$ with $a_{i}R_{3,n}d$ and $b_{i}R_{3,n}d$ for all $a_{i},b_{i}$. Also set $dR_{3,n}d$.
Then $\F_{3,n}$ validates $L$. Let $\M_{3,n}$ be defined as in Case 2 and let $d$ satisfy $r$ and any type in $T_{n}$. We have $\M_{3,n},a_{i}\models I_{n}$ for all $i<2^{n}$. Hence we have $\M_{3,n},a_{i}\models\chi_{n}$.
Pick for every $\xi\in \text{sub}(\chi_{n})$ satisfied in some world in $N_{2,n}$ a world $b_{\xi}$ in $N_{2,n}$ satisfying $\xi$ in $\M_{3,n}$. Also pick for every $\xi\in \text{sub}(\chi_{n})$ satisfied in some world in $M_{2,n}$ a world $a_{\xi}$ in $M_{2,n}$ satisfying $\xi$ in $\M_{3,n}$. Let $\M_{i}$ denote the restriction of $\M_{3,n}$ to 
\[
\{a_{i}\}\cup \{d\} \cup \{ b_{\xi}\mid b_{\xi} \in \text{sub}(\chi_{n}) \text{ picked}\} \cup \{ a_{\xi}\mid a_{\xi} \in \text{sub}(\chi_{n}) \text{ picked}\}.
\]
Now the proof continues similar to Cases 1 and 2: clearly each $\M_{i},c$ with $c$ in $\M_{i}$ satisfies the same type $t\in T_{n}$ as $\M_{3,n},c$ and it is easy to see that $\M_{i},a_{i}\models \chi_{n}$ for all $i<2^{n}$.
One can proceed in the same way as in the proof of Cases 1 and 2 and derive a contradiction.

\medskip

Case 4. $L \subseteq Log(\mathcal{F}_{2})$. Assume for a proof by contradiction that $\chi_{n}$ is equivalent to $I_{n}$ in $L$ and
of size $<2^{n}$. 

We construct a model $\M_{4,n}$ based on $\F_{4,n}=(W_{4,n},R_{4,n})$ as follows.
Define $W_{4,n}= M_{4,n}\cup N_{4,n}$ with $M_{4,n}= \{a_{0},\ldots,a_{2^{n}-1}\}$ and $N_{4,n}=\{b_{0},\ldots,b_{2^n-1}\}$ and let $R_{4,n}$ be the reflexive closure of $M_{4,n}\times N_{4,n}$.
Then $\F_{4,n}$ validates $L$. Let the valuation of $\M_{4,n}$ be defined in the same way as the valuation of $\M_{1,n}$ in Case~1. We have $\M_{4,n},a_{i}\models I_{n}$ for all $i<2^{n}$. Hence we have $\M_{4,n},a_{i}\models\chi_{n}$. Now one can proceed similar to the previous cases and derive a contradiction.

\medskip

Case 5. $L \subseteq Log(\mathcal{F}_{3})$. Assume for a proof by contradiction that $\chi_{n}$ is equivalent to $I_{n}$ in $L$ and
of size $<2^{n}$. 

We construct a model $\M_{5,n}$ based on $\F_{5,n}=(W_{5,n},R_{5,n})$ by adding to $\M_{4,n}$ a world $d$ with $a_{i}R_{5,n}d$ and $b_{i}R_{5,n}d$ for all $a_{i},b_{i}$. Also set $dR_{5,n}d$.
Then $\F_{5,n}$ validates $L$. Let the valuation of $\M_{5,n}$ be defined as in Case 4 and let $d$ satisfy $r$ and any type in $T_{n}$. We have $\M_{5,n},a_{i}\models I_{n}$ for all $i<2^{n}$. Hence we have $\M_{5,n},a_{i}\models\chi_{n}$. Now one can proceed similar to the previous cases and derive a contradiction.

\medskip

We continue with the pre-tabular normal modal logics containing GL.

\medskip

Case 6. $L \subseteq Log(\mathcal{F}_{\text{GL}.3})$. Assume for a proof by contradiction that $\chi_{n}$ is equivalent to $I_{n}$ in $L$ and
of size $<2^{n}$.

This case is essentially the same as Case 1 except that we have irreflexive instead of reflexive worlds.

\medskip

Case 7. $L \subseteq Log(\mathcal{F}_{n_{1},n_{2}})$. Assume for a proof by contradiction that $\chi_{n}$ is equivalent to $I_{n}$ in $L$ and
of size $<2^{n}$. This case is again similar to the previous cases.

Fix $n_{1}\geq 1$ and $n_{2}\geq 0$ and let $n> n_{1}+n_{2}$. We construct a model $\M_{7,n}$ based on $\F_{7,n}=(W_{7,n},R_{7,n})$ as follows:
let $W_{7,n}= M_{7,n}\cup N_{7,n}\cup P_{7,n}$ with $M_{7,n}= \{a_{0},\ldots,a_{2^{n}-1}\}$,
$N_{7,n}= \{b_{0},\ldots,b_{2^n-1}\}$ and $P_{7,n}=\{d_{1},\ldots d_{n_{1}},d_{n_{1}+1},\ldots,d_{n_{1}+n_{2}+1}\}$.
Define $R_{7,n}$ as the transitive closure of 
\[
(M_{7,n} \times (N_{7,n}\cup P_{7,n})) \cup \{(d_{i},d_{i+1})\mid 1\leq i<n_{1}+n_{2}\} \cup 
(N_{7,n}\times \{d_{n_{1}+1}\})
\]
Then $\F_{7,n}$ validates $L$. Let $\M_{7,n}$ be defined in such a way that
\begin{itemize}
    \item every type $t\in T_{n}$ is satisfied in exactly one $a_{i}$, $i<2^{n}$;
    \item $r$ is satisfied in $a_{i}$ for all $i<2^{n}$ and in all worlds in $P_{7,n}$;
    \item every type $t\in T_{n}$ is satisfied in exactly one $b_{i}$, $i<2^{n}$;
    \item the worlds in $P_{7,n}$ all satisfy one fixed type $t_{0}$.
\end{itemize}
We have $\M_{7,n},a_{i}\models I_{n}$ for all $i<2^{n}$. Hence we have $\M_{7,n},a_{i}\models\chi_{n}$.
Pick for every $\xi\in \text{sub}(\chi_{n})$ satisfied in some world in $N_{7,n}$ a world $b_{\xi}$ in $N_{7,n}$ satisfying $\xi$ in $\M_{7,n}$. Let $\M_{i}$ denote the restriction of $\M_{7,n}$ to 
\[
\{a_{i}\}\cup P_{7,n} \cup \{ b_{\xi}\mid b_{\xi} \in \text{sub}(\chi_{n}) \text{ picked}\}.
\]
Then each pointed $\M_{i},c$ with $c$ in $\M_{i}$ satisfies the same type $t\in T_{n}$ as $\M_{7,n},c$. Moreover, $\M_{i},a_{i}\models \chi_{n}$ for all $i<2^{n}$ can be shown by induction. Hence $\M_{i},a_{i}\models I_{n}$ since the $\M_{i},a_{i}$ are based on frames validating $L$. However, note that there is a type in $T_{n}$ that is not satisfied in any $b_{\xi}$ since the number of $b_{\xi}$ in $\M_{i}$ is $<2^{n}$. Hence there is an $a_{i}$ with $\M_{i},a_{i}\not\models I_{n}$ and we have derived a contradiction.
\end{proof}
\section{Proofs and Refutations of CIP}
\label{sec:smallproofs}
We show claims in Figure~\ref{fig:landscape} for which we have not found direct proofs in the literature.

\begin{theorem}\label{thm:altcip} Alt$_{1}$ and Alt$_{2}$ have UIP, whereas Alt$_{n}$ for $n\geq 3$ does not have CIP.
\end{theorem}
\begin{proof}
   CIP for Alt$_{1}$ is shown in~\cite{gabbay2006craig}. CIP for Alt$_{2}$ was proved in~\cite{DBLP:conf/stacs/JungK25} (where this logic is referred to as the logic of $\mathbb{T}_2$).
   CIP for Alt$_{n}$, $n\geq 3$ can be refuted using the same argument as in Example~\ref{example:first}. The results for UIP follow from the fact that a normal modal logic with CIP axiomatised by modal formulas of modal depth at most one also has UIP~\cite{DBLP:journals/corr/abs-2205-00448}.
\end{proof}
\begin{theorem}\label{thm:examplestab}
    EQ$_{1}$ and EQ$_{2}$ have UIP, whereas EQ$_{n}$ for $n\geq 3$ does not have CIP. The logics LO$_{n}$ for $n\geq 1$ have UIP.
\end{theorem}
\begin{proof}
    It is easy to see that, for $n\leq 2$, whenever two pointed EQ$_n$-models are bisimilar, they are isomorphic. Likewise for LO$_n$ ($n\geq 1$). It follows by 
    Theorem~\ref{thm:bisimCIP} that these logics have CIP, and hence UIP (recall that every tabular quasi-normal modal logic with CIP has UIP). Thus, these logics have UIP.

    On the other hand, let $n\geq 3$ and consider the EQ$_n$-models
    $M=(\{0,1,2\},\{0,1,2\}^2,V)$ and $N=(\{0,1,2\},\{0,1,2\}^2,V')$ with
        $V(p)=\{1\}$ and $V'(p)=\{1,2\}$.
    The pointed models $M,0$ and $N,0$ are bisimilar but not isomorphic. It follows
        by Theorem~\ref{thm:bisimCIP} that EQ$_n$ lacks CIP, and hence, lacks UIP.
\end{proof}

\end{document}